\documentclass[11pt]{article}

\usepackage[T1]{fontenc}
\usepackage[utf8]{inputenc}
\usepackage{lmodern}
\usepackage{microtype}
\usepackage[a4paper,margin=27mm]{geometry}
\usepackage{amsmath,amssymb,amsthm,mathtools,bm}
\usepackage{booktabs}
\usepackage{graphicx}
\usepackage{enumitem}
\usepackage[hidelinks]{hyperref}
\usepackage{url}
\usepackage{authblk}
\usepackage{float}

\allowdisplaybreaks
\usepackage{tikz}
\usetikzlibrary{arrows.meta, positioning}

\newtheorem{theorem}{Theorem}[section]
\newtheorem{proposition}[theorem]{Proposition}
\newtheorem{lemma}[theorem]{Lemma} 
\newtheorem{corollary}[theorem]{Corollary}
\theoremstyle{definition}
\newtheorem{definition}[theorem]{Definition}
\theoremstyle{remark}
\newtheorem{remark}[theorem]{Remark}

\newcommand{\A}{\mathcal A}
\newcommand{\Hs}{\mathcal H}
\newcommand{\Pp}{\mathcal P}
\newcommand{\Ss}{\mathcal Sh}
\newcommand{\Perm}{\mathcal S}
\newcommand{\C}{\mathcal C}
\newcommand{\Tr}{\operatorname{Tr}}
\newcommand{\diag}{\operatorname{diag}}
\newcommand{\supp}{\operatorname{supp}}
\newcommand{\comp}{\operatorname{comp}}
\newcommand{\rank}{\operatorname{rank}}
\newcommand{\spann}{\operatorname{span}}
\newcommand{\Stab}{\operatorname{Stab}}
\newcommand{\wt}{\operatorname{wt}}
\newcommand{\BHL}{\mathrm{BHL}}
\newcommand{\MC}{\mathrm{MC}}
\newcommand{\unr}{\mathrm{unr}}
\newcommand{\UC}{B^{\mathrm{UC}}_q}
\newcommand{\pe}{p_{\mathrm e}}
\newcommand{\eps}{\varepsilon}
\newcommand{\1}{\mathbf 1}

\usepackage{comment}
\usepackage{xcolor}

\title{Quantum channel bounds for constant-composition codes}

\author[1]{Artur Akhiiarov}
\author[2]{Peter Boyvalenkov}
\author[2]{Danila Cherkashin}
\author[3, 4, 5]{Andrei Raigorodskii}

\affil[1]{Laboratory of Advanced
Combinatorics and Network Applications,
Moscow Institute of Physics and Technology (State University),
Moscow, Russia}
\affil[2]{Institute of Mathematics and Informatics, Bulgarian Academy of Sciences, Sofia}
\affil[3]{Department of Discrete Mathematics and Laboratory of Advanced
Combinatorics and Network Applications,
Moscow Institute of Physics and Technology (State University),
Moscow, Russia}

\affil[4]{Department of Mathematical Statistics and Random Processes,
Faculty of Mechanics and Mathematics, Lomonosov Moscow State University,
Moscow, Russia}

\affil[5]{Caucasus Mathematical Center, Adyghe State University,
Maykop, Republic of Adygea, Russia}

\date{November 2026}

\begin{document}
\maketitle

\begin{abstract}
We extend the quantum-channel optimization of Alrabiah and Guruswami
to $q$-ary constant-composition codes.
We study explicit channel and flag constructions that recover selected classical
asymptotic upper bounds, including the Ben-Haim--Litsyn, moving-subspace
and Horn-channel bounds.
We give two equivalent formulations, Gram and flag, and prove stability of the obtained upper bounds under
permutation transport and generalized shortening.
A hierarchy defined by bounding the output-state ranks within each
flag improves strictly between consecutive levels at every composition and every positive
distance below the Plotkin threshold.

We provide finite matrix parametrizations of these bounds (i.e., the number of variables does not depend on the length of codes under consideration).
Numerical optimization rapidly produces improvements over the
known asymptotic upper bounds throughout all tested nonbinary
parameter ranges.
For rational input data, each finite level admits near-optimal
feasible rational descriptions of polynomial length.
Their feasibility can be checked exactly and their objective values
computed to any prescribed accuracy in polynomial time;
the associated threshold problem with a prescribed gap belongs
to $\mathsf{PromiseNP}$.


\end{abstract}

\section{Introduction}

Let $A_q(n,d)$ be the largest size of a $q$-ary code of length $n$ and minimum
Hamming distance $d$. Determining the asymptotic rate
$R_q(\delta)=\limsup_n\frac1n\log_qA_q(n,\lceil\delta n\rceil)$ is a central
open problem of coding theory; throughout, the relative distance $\delta$ belongs to $[0,1]$. The same question is asked for constant-weight $q$-ary
codes and, more generally, for constant-composition $q$-ary codes, in which every
codeword contains each alphabet symbol a prescribed number of times.

Upper bounds on $R_q(\delta)$ are traditionally obtained by bounding codes
inside a smaller space and transferring the estimate to the whole space, as in
the arguments of Elias, Bassalygo and Levenshtein~\cite{Eli55,Bas65,Lev71}, or by Delsarte linear programming~\cite{MRRW77,Samorodnitsky2001,BargJaffe2001}, or by combining these methods (see also \cite{Del1,Del2,Del3,DL,Lev95}).
Recent binary developments comprise OpenAI's moving-subspace
bounds~\cite{OAI26}, Alrabiah--Guruswami's mixed-qubit channel (MQC) and masked mixed-qubit channel (2MQC)
bounds~\cite{AG26}, Gay--Jeronimo--Liu's explicit Honeycomb and Horn-channel hierarchy bounds~\cite{GJL26}, and Salmon's
best-possible binary linear-programming bound~\cite{Sal26,salmon2026rate};
Barg~\cite{Bar26} commented on their common mechanism.

Alrabiah and Guruswami~\cite{AG26} introduced the \emph{pretty good criterion}:
if a suitable classical--quantum channel has pretty-good-measurement (PGM)
symbol error probability~\eqref{eq:SymbolError} below $\delta$ at the uniform composition, then its Holevo information~\eqref{eq:HolevoInf}
bounds the asymptotic code rate. Their $q$-ary criterion applies to
$\mathbb Z_q$-covariant channels~\cite[Proposition~27]{AG26}. In
\cite[\S7]{AG26} they formulate the optimization over all such channels,
denoted $R^\star_{q,\mathrm{cq}}(\delta)$, and introduce an explicit $q$-ary
mixed-channel ($q$-MC) family. The explicit $q$-MC bound is obtained by restricting the
optimization defining $R^\star_{q,\mathrm{cq}}(\delta)$ to
the $q$-MC family. The present paper studies the full optimization and its
extension to prescribed compositions.

For constant-weight and constant-composition codes, Akhiiarov,
Boyvalenkov, Cherkashin and Raigorodskii~\cite{ABCR26} (hereafter ABCR)
recently developed a permutation-transport framework that moves bounds between
composition rates at an exactly computed information cost, surveyed the available
binary and nonbinary bounds, and combined selected leading explicit bounds
into an optimized envelope $E_q(\delta,\omega)$, where $\omega \in [0,1]$ is the weight rate. 
The binary inputs for transport cover the moving-subspace, MQC and 2MQC bounds, the Honeycomb branch of the
Horn-channel hierarchy, and Salmon's bound all mentioned above; the nonbinary inputs include
Ben-Haim--Litsyn~\cite{BHL07} and the explicit $q$-MC bound of Alrabiah and Guruswami.
Furthermore, the transport operator of~\cite{ABCR26} improves some of these inputs. We use ABCR as the
reference for the comparisons between them.

\paragraph{Contribution.} In this paper we consider the following upper bound for constant-composition codes (Theorem~\ref{thm:main}).
For a composition rate $r = (r_0,\dots,r_{q-1})$, put
\[
 \UC(\delta;r) := \inf\bigl\{\chi_q(r,W):\ \pe(r,W)<\delta\bigr\},
\]
where $W$ ranges over all finite-dimensional classical--quantum channels, $\delta \in (0,1)$, $\chi_q(r,W)$ is Holevo information~\eqref{eq:HolevoInf}, and $\pe(r,W)$ is the PGM symbol error probability~\eqref{eq:SymbolError}; here $UC$ abbreviates universal channel.
This is the composition-dependent extension studied below. Its uniform
specialization is exactly the Alrabiah--Guruswami optimization:
\[
 U^{\mathrm{UC}}_q(\delta) := \UC \left(\delta; \left(\frac{1}{q},\dots,\frac{1}{q}\right)\right) = R^\star_{q,\mathrm{cq}}(\delta).
\]
Indeed, we shall see that in the uniform case a flagged cyclic symmetrization preserves both the PGM error and
the Holevo information at the uniform composition. Thus allowing arbitrary
channels does not change that optimized value. Our results concern the
structure, closure properties and comparisons of this channel bound, as
well as its extension to arbitrary compositions, explicit matrix optimization forms and numerics:

\begin{enumerate}[leftmargin=2em]
\item \emph{Structural formulation} (Theorem~\ref{thm:main}). We give
equivalent block Gram matrix and flag formulations of the channel bound
and prove that the function $H_q(r)-\UC(\delta;r)$, where $H_q$ is the $q$-ary entropy, is jointly concave in $(r,\delta)$.
The underlying Alrabiah--Guruswami criterion straightforwardly extends to every composition
$r$ and every finite-dimensional channel, so $\UC(\delta;r)$ is an upper bound for the code rate.
Of course, the uniform and symmetric compositions give the unrestricted and
constant-weight bounds (Corollaries~\ref{cor:unrestricted} and~\ref{cor:cw}).

\item \emph{Strict finite-level improvement} (Theorem~\ref{thm:strict}).
Let $B_q^{(m)}(\delta;r)$ denote the flagged bound with output state ranks
at most $m$ inside each flag, and let $F_q(\delta;r)$ denote the flagged $q$-ary moving-subspace (qMS) bound introduced in Section~\ref{sec:special}, see~\eqref{eq:flagged-qms}.
Denote the Plotkin threshold~\eqref{eq:PlotkinTreshold} by $\delta^\mathrm{Pl}_q(r)$.
For every $q\ge2$, every composition $r$, and
$0<\delta<\delta^\mathrm{Pl}_q(r)$, we prove
\[
 B_q^{(m+1)}(\delta;r)<B_q^{(m)}(\delta;r)
 \quad(m\ge1),
 \qquad
 B_q^{(q)}(\delta;r)<F_q(\delta;r).
\]
Consequently, every finite level lies strictly above the limit bound $\UC(\delta;r)$.
Together with the realizations of ABCR transport closure envelope $E_q(\delta,\omega)$ from~\cite{ABCR26}, for $r_\omega=(1-\omega,\tfrac{\omega}{q-1},\dots,\tfrac{\omega}{q-1})$ this gives
\[
 \UC(\delta;r_\omega)
 < B_q^{(q)}(\delta;r_\omega)
 < F_q(\delta;r_\omega)
 \leq E_q(\delta,\omega)
\]
for $q\ge3$, $0<\omega<1$, and
$0<\delta<\delta^\mathrm{Pl}_q(r_\omega)$, including the case of uniform composition
$\omega=(q-1)/q$ (Corollary~\ref{cor:envelope}).

\item \emph{Closure} (Section~\ref{sec:closure}). Permutation transport of~\cite{ABCR26} and generalized shortening of Laihonen and Litsyn~\cite{LL98} improve neither $\UC(\delta;r)$ nor any particular level $B_q^{(m)}(\delta;r)$.
Actually, we show that these methods are incorporated into our optimization (Theorem~\ref{thm:horn-closure} and Corollary~\ref{cor:universal}). 
By classification obtained in~\cite[Theorem 4.4]{ABCR26} this means, in particular, that the functions $B_q^{(m)}(\delta;r)$ and $\UC(\delta;r)$ are Schur concave (Corollary~\ref{cor:SchurConcavity}).

\item \emph{Comparison with explicit bounds} (Section~\ref{sec:special}). 
 The explicit $q$-MC channels of~\cite{AG26}, a larger covariant family, and the channels used in the Horn-channel
 hierarchy all belong to the class optimized over in the definition of $\UC(\delta;r)$.
The Ben-Haim--Litsyn bound~\cite{BHL07}, in the form used
in~\cite[\S5.4]{ABCR26}, combines two
single-component variational expressions associated with
two groups of code coordinates with prescribed relative weights.
For each feasible choice of the single-component parameters,
we construct a pure-state channel with matching Holevo information
and PGM error (Theorem~\ref{thm:bhl}).
A two-valued classical flag realizes the two-component
combination, including its entropy correction.
These constructions and the closure properties show that $\UC(\delta;r_\omega)$ lies below every
component of the ABCR envelope $E_q(\delta,\omega)$ (Subsection~\ref{sec:envelope}). For $q=2$
the optimization contains the MQC and 2MQC bounds of~\cite{AG26}, the
moving-subspace bounds of~\cite{OAI26}, the Honeycomb bound and Horn-channel
hierarchy of~\cite{GJL26}.
Note that the exact value of the Delsarte exponent in the binary unrestricted case was found by Salmon~\cite{Sal26,salmon2026rate} (see Subsection~\ref{sec:comparison-summary}).

\item \emph{Numerical evaluation and certification}  
(Sections~\ref{sec:computation}--\ref{sec:numerics}).
We give explicit finite-dimensional matrix parametrizations
for the flagged qMS bound and every level $B_q^{(m)}$,
and compare their numerical values with selected bounds
from the literature.
For rational input data, we prove that every finite level
admits arbitrarily close approximations to its optimum
by rational matrices and weights satisfying the composition
and error constraints exactly.
These certificates have polynomial description length,
and the associated threshold problem with a prescribed gap
belongs to $\mathsf{PromiseNP}$
(Section~\ref{sec:certificates}).
This provides a way to certify numerical upper bounds
independently of the optimization procedure.

\end{enumerate}

The proof of Theorem~\ref{thm:main} follows the argument of
Alrabiah and Guruswami~\cite{AG26}: the expected Hamming distance between a
codeword and its PGM estimate is controlled symbol by symbol, and a strong
converse turns a positive success probability into a rate bound. To treat
arbitrary compositions and arbitrary channels, we use Winter's
constant-composition strong converse~\cite{Win99} and concavity of the PGM
error in the composition. The latter replaces the output symmetry
assumption in the code-to-channel argument. 


The permutation ideas behind the transport inequalities of Elias, Bassalygo
and Levenshtein~\cite{Eli55,Bas65,Lev71} and their exact
constant-composition form in ABCR~\cite{ABCR26} enter through symmetrization
over groups of symbol permutations (Lemma~\ref{lem:sym}) and through the
flag construction (Lemma~\ref{lem:flags}). The flag construction realizes a
permutation coupling inside a single channel. Together with the Gram
formulation, it supplies the convexity and closure properties used to study
the full Alrabiah--Guruswami optimization and its composition-dependent
extension.

\paragraph{Organization.}
Section~\ref{sec:prelim} introduces the notation and the required
information-theoretic tools. We briefly describe the quantum--channel approach and the bound via PGM.
Section~\ref{sec:main} states the main theorem (constant-composition channel bound
and its structural properties), proved in Section~\ref{sec:proof}.
Section~\ref{sec:strict} introduces the rank-constrained hierarchy of the main bound and establishes strict improvement between
consecutive rank levels.
Section~\ref{sec:closure} proves stability of all our bounds under permutation transport
and generalized shortening and derives its consequences.
Section~\ref{sec:special} constructs the channel families that certify
the comparison with existing bounds.
Section~\ref{sec:computation} gives explicit finite matrix formulas of our level bounds.
Section~\ref{sec:certificates} proves the existence of short rational
certificates and $\mathsf{PromiseNP}$ inclusion.
Section~\ref{sec:numerics} presents numerical refinements.
Appendix~\ref{app:rays} studies pure-state perturbations and
behaviour near the Plotkin boundary.
Appendix~\ref{app:merging} treats symbol merging, and Appendix~\ref{app:shortening-fixed-points} characterizes the
fixed points of generalized shortening.

\section{Preliminaries}
\label{sec:prelim}

\subsection{Codes, compositions and rates}

The alphabet is $\A=[q]=\{0,1,\dots,q-1\}$, $q\ge2$. 
The \emph{weight} $\wt(x)$ of $x\in\A^n$ is its number of nonzero coordinates, and
$d_H(x,y)$ is the Hamming distance. The composition vector of a word $x=(x_1,\dots,x_n)$ is
$(n_0,\dots,n_{q-1})$, where $n_a=|\{j:x_j=a\}|$. The composition rate of $x$ is
$(n_0/n,\dots,n_{q-1}/n)$ (usually we abbreviate it to composition). A code is constant-composition if all its codewords have
the same composition vector. Following~\cite{ABCR26}, for a composition
vector $\bm w = (w_0,\dots,w_{q-1})$ with $\sum_a w_a=n$ let
\begin{align*}
 A_q(n,d)&=\max\{|C|:\ C\subseteq\A^n,\ d_H(x,y)\ge d\ (x\ne y\in C)\},\\
 A_q(n,d,w)&=\max\{|C|:\ \text{as above, }\wt(x)=w\ (x\in C)\},\\
 A_q(n,d;\bm w)&=\max\{|C|:\ \text{as above, every }x\in C\text{ has composition }\bm w\}.
\end{align*}

General composition rate vectors will be denoted by $r=(r_x)_{x \in \mathcal{A}}$, while the uniform and symmetric composition cases are given by
\[ 
u_q=\Bigl(\tfrac1q,\dots,\tfrac1q\Bigr),\qquad
 r_\omega=\Bigl(1-\omega,\tfrac{\omega}{q-1},\dots,\tfrac{\omega}{q-1}\Bigr),
\]
respectively.

Let 
\[
H_q(r)=-\sum_xr_x\log_qr_x, \qquad h_q(\omega)=H_q(r_\omega)
\]
be the $q$-ary entropy and its specialization to symmetric composition. All logarithms are to base $q$ unless written $\ln$. 
Note that $u_q=r_{\theta_q}$ with $\theta_q=(q-1)/q$.

The quantity 
\begin{equation} \label{eq:PlotkinTreshold}
 \delta^\mathrm{Pl}_q(r) :=1-\sum_xr_x^2 
\end{equation}
is the average distance between two independent symbols drawn from $r$; by the
Plotkin argument, constant-composition codes of composition rate $r$ and relative
distance $\delta>\delta^\mathrm{Pl}_q(r)$ have rate zero~\cite[\S5.8]{ABCR26}.

Let $\Delta_{q-1}$ be the probability simplex on $\A$. A vector
$r\in\Delta_{q-1}$ is an \emph{$n$-type} if $nr$ is integral.

\begin{definition}[Rates]
For $\delta>0$, $0 \leq \omega \leq 1$ and $r\in\Delta_{q-1}$,
\begin{gather*}
 R_q(\delta)=\limsup_{n\to\infty}\tfrac1n\log_qA_q(n,\lceil\delta n\rceil),\qquad
 R_q(\delta,\omega)=\limsup_{n\to\infty}\tfrac1n\log_qA_q(n,\lceil\delta n\rceil,\lceil\omega n\rceil),\\
 R_q(\delta;r)=\sup\Bigl\{\limsup_{j\to\infty}\tfrac1{n_j}\log_qA_q(n_j,\lceil\delta n_j\rceil;n_jr_j)\Bigr\},
\end{gather*}
where the supremum is over all sequences $n_j\to\infty$ and $n_j$-types
$r_j\to r$.
\end{definition}

For rational $r$, ABCR define $R_q(\delta;r)$ as the $\limsup$ along the
lengths $n$ with $nr$ integral~\cite[\S2]{ABCR26}; this is the special case
$r_j=r$ above, so every upper bound on our $R_q(\delta;r)$ bounds theirs. The
three rates are related as follows.

\begin{proposition}[{\cite[Proposition~4.2, Corollary 4.6]{ABCR26}}]
\label{prop:abcr-equalities}
For $q\ge2$, $\delta>0$ and $0 \leq \omega \leq 1$,
\[
 R_q(\delta)=R_q(\delta;u_q),\qquad
 R_q(\delta,\omega)=R_q(\delta;r_\omega).
\]
\end{proposition}

\subsection{Classical--quantum channels and the pretty good measurement}

In this subsection we give the PGM framework needed for the
quantum--channel approach following~\cite{AG26,dalai2017constant}.
The proof of Theorem~\ref{thm:main} retains the two PGM tools 
\cite[\S3 and \S7]{AG26} called coarse-graining (Lemma~\ref{lem:coarse}) and data processing (Lemma~\ref{lem:dp}). 
For the converse step at a prescribed composition,
we use Winter's constant-composition strong converse~\cite{Win99} (Theorem~\ref{thm:winter}) 
again following the approach of~\cite{AG26}.

A \emph{state} is a positive semidefinite matrix $\rho\succeq0$ with
$\Tr\rho=1$. A (finite-dimensional) \emph{classical--quantum channel}
$W:x\mapsto\rho_x$ assigns a state $\rho_x \in \mathbb C^{D\times D}$ to each input
symbol $x\in\A$; $D$ is the output dimension of $W$. 
A state $\rho$ is called \emph{pure} if
$\rank \rho=1$, or equivalently if
$\rho = \psi\psi^*$ for some unit vector $\psi$ (by $^*$ we denote the conjugate transpose).
A state that is not pure is called \emph{mixed}.
A classical channel is the special case of mutually commuting (simultaneously diagonalizable) states. 
Let a group $\mathfrak G$ act on $\A$, and let $g\mapsto U_g$ be a
unitary representation of $\mathfrak G$ on the output space $\mathbb C^D$.
A classical--quantum channel $W:x\mapsto\rho_x$ is
$\mathfrak{G}$-\emph{covariant} if
\[
 \rho_{g\cdot x}=U_g\rho_xU_g^* \qquad
 \text{for every } g\in \mathfrak G,\ x\in\A. 
\]

For an input
distribution $r$ write $\rho_r=\sum_xr_x\rho_x$ and define the von Neumann entropy as
\[
 S_q(\sigma)=-\Tr \sigma\log_q\sigma
\]
for a state $\sigma$ (with $0\log0=0$; we use this definition also for $\sigma\succeq0$ of any
trace). For a state $\sigma$, this is the entropy of its eigenvalues,
counted with multiplicity.
The \emph{Holevo information}~\cite{Hol73}, in base-$q$ units, is
\begin{equation} \label{eq:HolevoInf}
 \chi_q(r,W)=S_q(\rho_r)-\sum_xr_xS_q(\rho_x). 
\end{equation}
For states $A,B$ with $\supp A\subseteq\supp B$, define \textit{quantum relative entropy}
\[
 D_q(A\Vert B) := \Tr A(\log_q A-\log_q B);
\]
otherwise set $D_q(A\Vert B)=+\infty$. For diagonal states $A=\diag r$ and $B=\diag s$,
this reduces to the classical relative entropy
\[
 D_q(r\Vert s)=\sum_{x:r_x>0}r_x\log_q\frac{r_x}{s_x},
\]
with value $+\infty$ if $r_x>0=s_x$ for some $x$.

A \emph{positive operator-valued measure} (POVM) is a finite family of
matrices $M_y\succeq0$ with $\sum_yM_y=I$. On a state $\rho$, outcome $y$
has probability $\Tr(\rho M_y)$. A \emph{sub-POVM} satisfies
$\sum_yM_y\preceq I$; the remaining (positive semidefinite) operator $I-\sum_yM_y$ represents
a failure outcome.

For a finite ensemble $\{(\pi_c,\sigma_c)\}_{c \in C}$ of probabilities $\pi_c$ and states $\sigma_c$, a decoding POVM or sub-POVM with operators $M_c$ has conditional error $e_c=1-\Tr(\sigma_cM_c)$, average error $\sum_c\pi_ce_c$, and maximal
error $\max_c e_c$.
We omit ensemble components of probability zero; the same convention
applies to conditional ensembles below. In the paper we use the ensembles based on the codes, so we use the same letter $C$ for the index set.

For an ensemble with average $\sigma=\sum_c\pi_c\sigma_c$, the
\emph{pretty good measurement} (PGM)~\cite{HW94} is the family of operators
\[
 M_c=\sigma^{-1/2}\pi_c\sigma_c\sigma^{-1/2},
\]
where $\sigma^{-1/2}$ is the inverse square root on
$\supp \sigma$ and is zero on its orthogonal complement.
These operators sum to the orthogonal projection onto
$\supp \sigma$, so they form a POVM on that support.
Its \emph{PGM error} is the average probability that the decoded label differs
from the transmitted label:
\begin{equation} \label{eq:SymbolError}
 \pe=1-\sum_c\pi_c\Tr(\sigma_cM_c), 
\end{equation}
see~\cite[\S2]{AG26}. For a channel $W$ and composition $r$ we
write $\pe(r,W)$ for the PGM error of the ensemble
$\{(r_x,\rho_x)\}_{x\in\A}$.

\begin{lemma}[PGM coarse-graining {\cite[Lemma~8]{AG26}}]
\label{lem:coarse}
Consider an ensemble $\{(\pi_c,\sigma_c)\}_{c\in C}$ with PGM
operators $M_c$. Let $\C$ be the transmitted label, with
$\Pr[\C=c]=\pi_c$, $c\in C$, and let $\widehat \C$ be the corresponding PGM outcome.
For a map $g: C \to \mathcal Y$, the operators
$\sum_{c:g(c)=y}M_c$ form the PGM of the push-forward ensemble
\[
 \bigl\{(\Pr[g(\C)=y],\ \mathbb E[\sigma_\C\mid g(\C)=y])\bigr\}_y.
\]
Consequently, $\Pr[g(\widehat \C)\ne g(\C)]$ is the PGM error of this
push-forward ensemble.
\end{lemma}

A linear map $\Phi$ between matrix spaces is \emph{completely positive}
if $\operatorname{id}_k\otimes\Phi$ preserves positive semidefiniteness
for every positive integer $k$, and \emph{trace-preserving} if
$\Tr \Phi(A)=\Tr A$ for every matrix $A$ in its domain.

\begin{lemma}[PGM data processing
{\cite[Lemma~10]{AG26},~\cite[\S11.5]{Ren22}}]
\label{lem:dp}
Applying a completely positive trace-preserving map to all states of an
ensemble does not decrease its PGM error.
\end{lemma}

\begin{theorem}[Constant-composition strong converse
{\cite[Theorem~13 and its proof]{Win99}}]
\label{thm:winter}
For every input alphabet size $q$, output dimension $D$ and
$\lambda\in(0,1)$, there is a constant $\kappa=\kappa(q,D,\lambda)$
with the following property. Let $W$ be a classical--quantum channel
with input alphabet $\A$ and output dimension $D$, and let
$c^{(1)},\dots,c^{(N)} \in \A^n$ be distinct codewords of a common
type $r$ with a decoding sub-POVM whose maximal error probability
on the product states $\sigma_{c^{(j)}}$ is at most $\lambda$, where
\[
 \sigma_c=\rho_{c_1}\otimes\cdots\otimes\rho_{c_n}
 \qquad\text{for }c=(c_1,\dots,c_n).
\]
Then
\begin{equation}\label{eq:keybound}
 \ln N
 \leq n\ln q\cdot\chi_q(r,W)
 +\kappa\sqrt n+\ln\frac4{1-\lambda}.
\end{equation}
\end{theorem}

The first display of the proof of~\cite[Theorem~13]{Win99} gives the bound~\eqref{eq:keybound} for codewords of the type $r$, with $\kappa=2KD\sqrt q\,\delta_W$,
$\delta_W=\sqrt{32qD}/(1-\lambda)$ and an absolute constant $K$;
in particular, one sees that $\kappa$ does not depend on $r$ or $W$.
In fact, we shall use the right-hand side of~\eqref{eq:keybound} for fixed $q$, $D$ and $\lambda$ as $n\ln q\cdot\chi_q(r,W) + O(\sqrt{n})$ with $n \to \infty$.

\section{The Alrabiah--Guruswami optimization at fixed composition}
\label{sec:main}

\subsection{Gram data}

The objective and error constraint in the channel optimization of
Alrabiah and Guruswami~\cite[\S7]{AG26} admit a block Gram formulation.
We develop such a formulation for arbitrary input compositions and mixed output states;
it will also make direct sums and permutation symmetrization explicit.
For nonnegative integers $d_x$, put
\[
 \Hs_x=\mathbb C^{d_x},\qquad
 \Hs=\bigoplus_{x\in\A}\Hs_x.
\]
We call $\Hs_x$ the $x$-th \emph{slot} and $\Hs$ the
\emph{slot space}. Let $J_x:\Hs_x\to\Hs$ denote the
canonical inclusion. A \emph{Gram datum} is a matrix
$G\succeq0$ on this specified slot space with $\Tr G=1$.
Its blocks are
\[
 G_{xy}=J_x^*GJ_y\in\mathbb C^{d_x\times d_y}.
\]
Note that $\dim\Hs=\sum_xd_x$; we call $\max_xd_x$ the \emph{level} of the specified Gram datum.
We set
\[
 \comp(G)_x=\Tr G_{xx},\qquad
 \phi(G)=S_q(G)-\sum_xS_q(G_{xx}),\qquad
 \pe(G)=1-\sum_x\Tr\bigl[\bigl((G^{1/2})_{xx}\bigr)^2\bigr].
\]
Thus $\comp(G)\in\Delta_{q-1}$ is the composition rate of $G$, and
$\phi(G)=-D_q(G\,\|\,\textstyle\bigoplus_xG_{xx})\leq 0$ is minus the relative
entropy of $G$ to its block-diagonal part. 
For a channel $W:x\mapsto\rho_x$ and a composition $r$, choose
factorizations $r_x\rho_x=R_xR_x^*$ and set
$R=[R_0\cdots R_{q-1}]$ and $G=R^*R$.
Lemma~\ref{lem:gram} below shows that $G$ is a Gram datum satisfying
$\comp(G)=r$,
$\chi_q(r,W)=H_q(r)+\phi(G)$ and $\pe(r,W)=\pe(G)$,
and that every Gram datum arises in this way.
For fixed slot spaces, the function $\phi$ is concave by joint convexity of relative entropy.
Minimizing this concave functional over fixed slot spaces
is generally a nonconvex problem.
Section~\ref{sec:certificates} nevertheless shows that
near-optimal finite-level bounds admit polynomial-size
rational certificates, with exact feasibility checks
and polynomial-time verification of the associated
threshold problem with a prescribed gap.
The convexity of the entropy-adjusted optimized value
comes from mixing flag families.
This preserves the rank bound within each component,
although representing the mixture by a single Gram datum
may require larger slots.

\subsection{Constant-composition extension and structural properties}

We extend the optimization $R^\star_{q,\mathrm{cq}}$ of
Alrabiah and Guruswami~\cite[\S7]{AG26} to a prescribed input
composition and allow all finite-dimensional classical--quantum channels. The notation
$\UC$ refers to this extension. Corollary~\ref{cor:unrestricted} below
identifies its uniform-composition value with the original value $R^\star_{q,\mathrm{cq}}(\delta)$.

\begin{definition}[Constant-composition channel bound]
For $r\in\Delta_{q-1}$ and $\delta>0$ let
\[
 \UC(\delta;r)=\inf\bigl\{\chi_q(r,W):\ \pe(r,W)<\delta\bigr\},
\]
the infimum over all classical--quantum channels with input alphabet $\A$
and arbitrary finite output dimension. We write
$\UC(\delta,\omega)=\UC(\delta;r_\omega)$ for a symmetric composition $r_\omega$ and
$U^{\mathrm{UC}}_q(\delta)=\UC(\delta;u_q)$ for the uniform composition $u_q$.
At zero distance we set $\UC(0;r)=H_q(r)$ and $U^{\mathrm{UC}}_q(0)=1$.
\end{definition}

These values agree with the right limits.
Let $X$ have distribution $r$, and let $\widehat X\in\A$
be the PGM outcome for the ensemble $\{(r_x,\rho_x)\}_x$.
Write $e=\Pr[\widehat X\ne X]=\pe(r,W)$.
Fano's inequality~\cite[Eq.~(2.140)]{CT06} gives
$H_q(X\mid\widehat X)\le h_q(e)$.
Consequently, the Holevo bound gives
\begin{equation}\label{eq:fano}
 H_q(r)-h_q(e)
 \le I_q(X;\widehat X)
 \le \chi_q(r,W),
\end{equation}
where $I_q$ denotes classical mutual information in base-$q$ units.

For $0<\delta\le\theta_q$, monotonicity of $h_q$ on
$[0,\theta_q]$ therefore gives
\[
 H_q(r)-h_q(\delta)\le\UC(\delta;r)\le H_q(r).
\]
The upper bound follows from the channel with orthogonal pure outputs, whose Holevo information is $H_q(r)$
and whose PGM error is zero.

We use $\|A\|_F$ for the Frobenius norm $\sqrt{\Tr(A^*A)}$ (also known as Hilbert--Schmidt); note that $A$ is not necessarily a square matrix.

\begin{theorem}[Constant-composition extension and structure]\label{thm:main}
Let $q\ge2$, $\delta>0$ and $r\in\Delta_{q-1}$. Then
\[
 R_q(\delta;r)\leq \UC(\delta;r).
\]
Moreover:
\begin{enumerate}[label=(\alph*),leftmargin=2em]
\item \emph{(Gram form)} one has
\begin{equation} \label{eq:mainThGramform}
\UC(\delta;r)=H_q(r)+\inf\{\phi(G):\ \comp(G)=r,\ \pe(G)<\delta\}, 
\end{equation}
where the infimum is over Gram data of all finite dimensions $d_x$. 
Equivalently,
$\UC(\delta;r)=H_q(r)-\sup\{D_q(G\,\|\,\bigoplus_xG_{xx})\}$ over Gram data with
$\comp(G)=r$ and $\sum_{x\ne y}\|(G^{1/2})_{xy}\|_F^2<\delta$.

\item \emph{(Flag form)} the value $\UC(\delta;r)$ equals the infimum of
\begin{equation} \label{eq:mainThFlagform}
 H_q(r)+\sum_{i \in I} w_i\bigl[\chi_q(v_i,W_i)-H_q(v_i)\bigr]
\end{equation}
over all finite families of weights $w_i\ge0$, $\sum_i w_i=1$, all compositions
$v_i\in\Delta_{q-1}$ with $\sum_iw_iv_i=r$ and all channels $W_i$ with
$\sum_iw_i\pe(v_i,W_i)<\delta$. 
Moreover, it is always possible to take $I_0 \subset I$ with $|I_0|\leq q+1$ and change the weights putting zeros for $I\setminus I_0$ and preserving the
composition and average error without increasing the objective.

\item \emph{(Structure)} The function
$(r,\delta)\mapsto H_q(r)-\UC(\delta;r)$
is jointly concave on $\Delta_{q-1}\times(0,1)$,
nondecreasing in $\delta$, and invariant under permutations
of $\A$. Further $0\leq \UC(\delta;r)\leq H_q(r)$, $\UC(\delta;r)=0$ for $\delta\ge\delta^\mathrm{Pl}_q(r)$
(if $\delta^\mathrm{Pl}_q(r)>0$), and for $0<\delta<1$ the constraint $<\delta$ may be
replaced by $\leq \delta$ in~\eqref{eq:mainThGramform} and~\eqref{eq:mainThFlagform} without changing
the value.
\end{enumerate}
\end{theorem}

Theorem~\ref{thm:main} is proved in Subsection~\ref{subsec:proofMainTh}.

In the flag form, only the \emph{average} error is constrained: a family of
channels, each used at its own composition $v_i$, is merged into one
channel whose output also reveals the index $i$ (a ``flag''). The term
$H_q(r)-\sum_iw_iH_q(v_i)$ is the mutual information between the input and
the flag, which is the information-theoretic cost of the flag.

\begin{corollary}[Unrestricted codes]\label{cor:unrestricted}
For $q\ge2$ and $\delta>0$,
\[
 R_q(\delta)\leq U^{\mathrm{UC}}_q(\delta)=1+\inf\{\phi(G):\ \pe(G)<\delta\},
\]
where the composition of $G$ is not restricted. Moreover
$U^{\mathrm{UC}}_q(\delta)$ equals the infimum of $\chi_q(u_q,W)$ over channels
that are $\mathbb Z_q$-covariant and satisfy $\pe(u_q,W)<\delta$.
\end{corollary}

The cyclic output symmetry used in~\cite[\S7]{AG26}
is equivalent to $\mathbb Z_q$-covariance. Indeed, if a
unitary $V$ cyclically permutes the output states, choose
a unitary $q$-th root $T$ of $V^q$ as follows. 
Write $V^q=\sum_\lambda\lambda P_\lambda$ and choose
$\mu_\lambda$ with $\mu_\lambda^q=\lambda$.
Set $T=\sum_\lambda\mu_\lambda P_\lambda$.
Then $T$ commutes with $V$ and every $\rho_x$, so
$V':=VT^{-1}$ satisfies $(V')^q=I$ and induces the same
permutation of the output states. Corollary~\ref{cor:unrestricted} (which is proved in Subsection~\ref{sec:proof-cor}) means that unrestricted value is exactly the optimization introduced by Alrabiah and Guruswami~\cite[\S7, Table~3]{AG26}:
\[
 U^{\mathrm{UC}}_q(\delta)
 =\UC(\delta;u_q)
 =R^\star_{q,\mathrm{cq}}(\delta).
\]
For every $\mathbb Z_q$-covariant channel $W$,
\cite[Proposition~27]{AG26} proves the rate bound $R_q(\delta)\leq \chi_q(u_q,W)$
provided that $\pe(u_q,W)<\delta$.
The additional assertion here is that
allowing arbitrary channels does not lower the infimum, and that the same
value has the unrestricted-composition Gram representation above.
More is known about the binary case.

\begin{corollary}[Binary specialization: Salmon's theorem~\cite{salmon2026rate}]
\label{cor:salmon}
Let $D_2(n,d)$ denote the optimum of the binary Delsarte
linear program. For every $0<\delta<1/2$,
\[
 U^{\mathrm{UC}}_2(\delta)
 = \lim_{n\to\infty}\frac1n
 \log_2 D_2(n,\lceil\delta n\rceil)
 = \mathcal J(\delta),
\]
where $\mathcal J$ is the explicit elliptic-integral function
in~\cite[Equation~(1.1)]{salmon2026rate}.
\end{corollary}

\begin{proof}
Corollary~\ref{cor:unrestricted} identifies
$U^{\mathrm{UC}}_2(\delta)$ with the infimum over finite-dimensional
binary output-symmetric channels. The conclusion follows from
\cite[Theorem~1.1 and Remark~1.2]{salmon2026rate}.
\end{proof}

\begin{corollary}[Constant-weight codes]\label{cor:cw}
For $q\ge2$, $0\leq \omega\leq 1$ and $\delta>0$,
\[
 R_q(\delta,\omega)\leq \UC(\delta,\omega)=h_q(\omega)+\inf\{\phi(G):\
 \Tr G_{00}=1-\omega,\ \pe(G)<\delta\}.
\]
\end{corollary}

Corollaries~\ref{cor:unrestricted} and~\ref{cor:cw}
also follow from Theorem~\ref{thm:main} and Proposition~\ref{prop:abcr-equalities}; the direct proofs in
Section~\ref{sec:proof-cor} additionally identify the
right-hand sides.

\section{Proof of the main theorem}
\label{sec:proof}

\subsection{Gram calculus}

We first rewrite the Holevo information and PGM error appearing in the
Alrabiah--Guruswami criterion~\cite{AG26} in terms of block Gram data.
The passage between ensembles and Gram matrices is standard
\cite[\S2]{JS00}; the direct-sum identities below provide the structural
properties needed for the optimization.

\begin{lemma}[Gram representation]\label{lem:gram}
Let $W:x\mapsto\rho_x$ be a channel with output dimension $D$ and
$r\in\Delta_{q-1}$. For $r_x>0$ factor $r_x\rho_x=R_xR_x^*$ with
$R_x:\mathbb C^{d_x}\to\mathbb C^D$ (one may take $d_x=0$ and $R_x=0$ if $r_x=0$). Let
$R=[R_0\cdots R_{q-1}]$ and $G=R^* R$. Then $G$ is a Gram datum with
$\comp(G)=r$ and Holevo information and PGM error are given by
\[
 \chi_q(r,W)=H_q(r)+\phi(G),\qquad \pe(r,W)=\pe(G).
\]
Conversely, every Gram datum $G$ arises in this way from a channel of output
dimension $\dim\Hs$ in which $\rho_x$ has rank at most $d_x$ for every symbol
of positive composition.
\end{lemma}

\begin{proof}
The matrices $\rho_r=RR^*$ and $G=R^* R$ have the same nonzero
eigenvalues, so $S_q(\rho_r)=S_q(G)$. Likewise $r_x\rho_x=R_xR_x^*$
and $G_{xx}=R_x^* R_x$ give
$S_q(G_{xx})=r_xS_q(\rho_x)-r_x\log_qr_x$; summing over $x$ yields
$\sum_xr_xS_q(\rho_x)=\sum_xS_q(G_{xx})-H_q(r)$, hence the formula for
$\chi_q$. For the error,
$1-\pe(r,W)=\sum_x\Tr\bigl[(R_x^*\rho_r^{-1/2}R_x)^2\bigr]$, and by the
singular value decomposition of $R$,
$R^*(RR^*)^{-1/2}R=(R^* R)^{1/2}$, whose $xx$ block is
$R_x^*\rho_r^{-1/2}R_x$. Conversely, given $G$ take $R=G^{1/2}$,
$R_x$ its block columns, $\rho_x=R_xR_x^*/\comp(G)_x$, and any fixed
state if $\comp(G)_x=0$. 
\end{proof}

For a fixed channel, the choice $R_x=\sqrt{r_x}\rho_x^{1/2}$ makes $G$ a
continuous function of $r$; hence the PGM error $\pe(r,W)$ and Holevo information $\chi_q(r,W)$ are continuous
in $r$.

\begin{lemma}[Direct sums and flags]\label{lem:flags}
For Gram data $G_i$ and weights $w_i\ge0$ with $\sum_i w_i=1$, the direct sum
$G=\bigoplus_iw_iG_i$, indexed by a finite set $I$, with slots $\bigoplus_i(\Hs_i)_x$, satisfies
\[
 \comp(G)=\sum_iw_i\comp(G_i),\qquad \pe(G)=\sum_i w_i\pe(G_i),\qquad
 \phi(G)=\sum_iw_i\phi(G_i).
\]
In channel language: for weights $w_i$, compositions $v_i$ with
$\sum_iw_iv_i=r$ and channels $W_i:x\mapsto\rho_{i,x}$, the \emph{flagged channel}
$W^{\mathrm{fl}}:x\mapsto\sigma_x$, where 
$\sigma_x=\bigoplus_i\frac{w_iv_i(x)}{r_x}\rho_{i,x}$ for $r_x>0$
(with an arbitrary state assigned when $r_x=0$) satisfies
\begin{equation}\label{eq:flag}
 \pe(r,W^{\mathrm{fl}})=\sum_iw_i\pe(v_i,W_i),\qquad
 \chi_q(r,W^{\mathrm{fl}})=H_q(r)+\sum_iw_i\bigl[\chi_q(v_i,W_i)-H_q(v_i)\bigr].
\end{equation}
\end{lemma}

\begin{proof}
Use the canonical identification
\[
 \bigoplus_i\Hs_i
 =\bigoplus_i\bigoplus_x(\Hs_i)_x
 \cong\bigoplus_x\Hs_x,
 \qquad
 \Hs_x:=\bigoplus_i(\Hs_i)_x.
\]
With respect to these slots,
\[
 G_{xx}=\bigoplus_iw_i(G_i)_{xx},
 \qquad
 (G^{1/2})_{xx}
 =\bigoplus_i\sqrt{w_i}(G_i^{1/2})_{xx}.
\]
Taking traces in the first identity gives the composition formula.
The second identity gives
\[
 \Tr\bigl((G^{1/2})_{xx}^2\bigr)
 =\sum_iw_i\Tr\bigl((G_i^{1/2})_{xx}^2\bigr),
\]
which proves the error formula. Further
$S_q(G)=\sum_{i\in I}w_iS_q(G_i)+H_q(w)$ and, since
$\sum_x\Tr(G_i)_{xx}=1$,
$\sum_xS_q(G_{xx})=\sum_i w_i \sum_xS_q((G_i)_{xx})+H_q(w)$. For each $i$, choose factors
$v_i(x)\rho_{i,x}=R_{i,x}R_{i,x}^*$ with $G_i=R_i^*R_i$.
The flagged channel at composition $r$ has factors
\[
 R_x=\bigoplus_i\sqrt{w_i}\,R_{i,x},
 \qquad r_x\sigma_x=R_xR_x^*.
\]
Its Gram datum is therefore $\bigoplus_iw_iG_i$.
Lemma~\ref{lem:gram}, applied to this channel at composition $r$, gives~\eqref{eq:flag}.
\end{proof}

\begin{lemma}[Relabelling and symmetrization]\label{lem:sym}
For a permutation $g$ of $\A$ let $G^g$ be $G$ with slot label $x$ renamed $g(x)$.
Then $\comp(G^g)=\comp(G)\circ g^{-1}$, while $\pe$ and $\phi$ are unchanged.
For a group $\mathfrak G \leq \Perm_q$ (where $\Perm_q$ is the symmetric group on $\A$) put
$G^{\mathfrak G}=\bigoplus_{g\in\mathfrak G}|\mathfrak G|^{-1}G^g$. Then
$\comp(G^{\mathfrak G})$ is the $\mathfrak G$-average of $\comp(G)$,
$\pe(G^{\mathfrak G})=\pe(G)$, $\phi(G^{\mathfrak G})=\phi(G)$, and the channel
realizing $G^{\mathfrak G}$ by Lemma~\ref{lem:gram} is
$\mathfrak G$-covariant (modulo states of weight zero).
\end{lemma}

\begin{proof}
The first part is clear and the second follows from Lemma~\ref{lem:flags}.
For $g\in\mathfrak G$, the coordinate permutation $U_g$ sending block $h$,
slot label $x$ to block $gh$, slot label $g(x)$ commutes with $G^{\mathfrak G}$, hence
with $R=(G^{\mathfrak G})^{1/2}$. Therefore
$R_{g(x)}R_{g(x)}^*=U_gR_xR_x^* U_g^*$ and
$\comp(G^{\mathfrak G})_{g(x)}=\comp(G^{\mathfrak G})_x$. (For symbols of
zero weight choose a common invariant state, such as the maximally mixed state.)
\end{proof}

If the $i$-th Gram datum has slot dimensions $d_{i,x}$,
their direct sum has slot dimensions $d_x=\sum_i d_{i,x}$.
Thus flags and symmetrization need not preserve a fixed
Gram level. This is allowed here because the optimization
ranges over all finite slot dimensions. Each resulting
channel is fixed before the limit $n\to\infty$ is taken.

Lemma~\ref{lem:sym} is the channel counterpart of the averaging
arguments of Bassalygo and Levenshtein~\cite{Bas65,Lev71}
and of the permutation transport of ABCR~\cite[Theorem~2.2]{ABCR26}.
Each coordinatewise isometry preserves Hamming distance.
The flagged channel construction preserves the PGM error
and the entropy offset $\phi$.
If $\bar r$ is the averaged composition and $\widetilde W$ the flagged
symmetrization, then
$\chi_q(\bar r,\widetilde W)=\chi_q(r,W)+H_q(\bar r)-H_q(r)$.

Alrabiah and Guruswami prove that the uniform composition maximizes the PGM
error for their output-symmetric channels~\cite[Lemma~11 and \S7]{AG26}.
The following concavity statement supplies the corresponding ingredient
for a fixed, possibly nonuniform composition.

\begin{lemma}[Concavity in the composition]\label{lem:concave}
For a fixed channel $W$, the maps $r\mapsto\pe(r,W)$ and
$r\mapsto\chi_q(r,W)$ are concave. If $W$ is $\mathfrak G$-covariant and
$\bar r$ is the $\mathfrak G$-average of $r$, then $\pe(r,W)\leq \pe(\bar r,W)$
and $\chi_q(r,W)\leq \chi_q(\bar r,W)$.
\end{lemma}

\begin{proof}
Let $r=\sum_j\alpha_jr^{(j)}$. For $r_x>0$ set $\hat\rho_x=\bigoplus_j\frac{\alpha_jr^{(j)}_x}{r_x}\rho_x$.
Let $\widehat W:x\mapsto\hat\rho_x$. By Lemma~\ref{lem:flags} with $w_j=\alpha_j$, $v_j=r^{(j)}$ and $W_j=W$, one has $\pe(r,\widehat W)=\sum_j\alpha_j\pe(r^{(j)},W)$. 
The partial trace over the flag register, defined on all
block matrices by $\Phi((\sigma_{jk})_{j,k})=\sum_j \sigma_{jj}$,
is completely positive and trace-preserving (see~\cite[Corollary~2.27]{Wat18}).
It sends $\hat\rho_x$ to $\rho_x$ whenever $r_x>0$, so
Lemma~\ref{lem:dp} gives
$\sum_j\alpha_j\pe(r^{(j)},W)\leq \pe(r,W)$. Concavity of $\chi_q(r,W)$ in $r$ follows from concavity of
the von Neumann entropy \cite[Theorem~5.23]{Wat18}, since
$r\mapsto\sum_xr_x\rho_x$ is affine and
$\sum_xr_xS_q(\rho_x)$ is linear. Covariance gives
$\pe(r\circ g^{-1},W)=\pe(r,W)$ and the same for $\chi_q$; average over
$\mathfrak G$.
\end{proof}

\subsection{From symbols to codes}

We follow the symbol-by-symbol PGM argument of Alrabiah and Guruswami
\cite[Theorem~17, Corollary~18 and Proposition~27]{AG26}. 
Our change is to average the coordinate input distributions,
whose mean is the common codeword composition, and use
Lemma~\ref{lem:concave}.

\begin{lemma}[Expected distance to the PGM estimate]\label{lem:distance}
Let $W:x\mapsto\rho_x$ be a channel, 
and let $C$ be a constant-composition code of composition rate $r$. 
Let $X$ be uniform on $C$, and let $\hat X\in C$ be the PGM outcome for the ensemble
$\{(1/|C|,\rho_{c_1}\otimes\dots\otimes\rho_{c_n})\}_{c\in C}$. Then
\[
\mathbb E\,d_H(X,\hat X)\leq n\,\pe(r,W).
\]
\end{lemma}

\begin{proof}
Let $r^{(i)}$ be the distribution of the coordinate $X_i$. Then
$\frac1n\sum_ir^{(i)}=r$, because every codeword has composition $nr$. By
Lemma~\ref{lem:coarse} with $g(c)=c_i$, $\Pr[\hat X_i\ne X_i]$ is the PGM
error of the ensemble 
\[
\{(r^{(i)}_x,\mathbb E[\rho_{X_1}\otimes\dots\otimes\rho_{X_n}\mid X_i=x])\}_{x\in \A}.
\]
The partial trace over all tensor factors except the $i$-th is the
linear map determined by
\[
 A_1\otimes\cdots\otimes A_n
 \longmapsto
 \left(\prod_{j\ne i}\Tr A_j\right)A_i.
\]
This map is completely positive and trace-preserving (see~\cite[Corollary~2.27]{Wat18}). It sends the
states in the preceding display to $\rho_x$, so Lemma~\ref{lem:dp} gives
$\Pr[\hat X_i\ne X_i]\leq \pe(r^{(i)},W)$. Summing over $i$ and using
Lemma~\ref{lem:concave} we have
\[
\mathbb E\,d_H(X,\hat X)\leq \sum_i\pe(r^{(i)},W)\leq n\,\pe(r,W).
\]
\end{proof}

\begin{proposition}[Constant-composition form of the pretty good criterion]\label{prop:criterion}
Fix a channel $W$ of output dimension $D$.
\begin{enumerate}[label=(\roman*),leftmargin=2em]
\item Let $C$ be a constant-composition code with composition vector $nr$, the minimum
distance at least $\delta n$, and
\[
 0< \varepsilon \leq 1-\pe(r,W)/\delta.
\]
Then
\[
 \log_q|C|\leq n\,\chi_q(r,W)+\frac{\kappa(q,D,1-\eps/2)}{\ln q}\sqrt n+\log_q\frac{16}{\eps^2}.
\]
\item If $\pe(r,W)<\delta$, then $R_q(\delta;r)\leq \chi_q(r,W)$.
\end{enumerate}
\end{proposition}

\begin{proof}
(i) A decoding error moves $\hat X$ to another codeword, at distance at least
$\delta n$; by Lemma~\ref{lem:distance}, $\Pr[\hat X\ne X]\leq \pe(r,W)/\delta$, so
the average success probability is at least $\eps$. The codewords with
individual success probability at least $\eps/2$ form a subcode $C'$ with
$|C'|\geq \eps |C|/2$. With the same measurement operators $C'$ has maximal error at most
$1-\eps/2$, and all its codewords have type $r$. Apply
Theorem~\ref{thm:winter} to $C'$ with $\lambda=1-\eps/2$ and use $\log_q |C| \leq \log_q |C'| + \log_q{\frac{2}{\varepsilon}}$.

(ii) Let $r_j\to r$ be $n_j$-types. By continuity,
$\pe(r_j,W)\leq \frac12(\pe(r,W)+\delta)<\delta$ for large $j$, and
$\chi_q(r_j,W)\to\chi_q(r,W)$. Apply (i) at $r_j$ with the fixed margin
$\eps=\frac12(1-\pe(r,W)/\delta)$ and divide by $n_j$.
\end{proof}

As in the pretty good criterion of Alrabiah and Guruswami~\cite{AG26},
the channel is fixed before $n\to\infty$; the strict inequality
$\pe(r,W)<\delta$ is what makes the success probability bounded away from
zero.

\subsection{Proof of Theorem~\ref{thm:main}} \label{subsec:proofMainTh}

\emph{The bound.} Let $G$ be a Gram datum with $\comp(G)=r$ and
$\pe(G)<\delta$. Lemma~\ref{lem:gram} gives a channel with $\pe(r,W)=\pe(G)$
and $\chi_q(r,W)=H_q(r)+\phi(G)$, and Proposition~\ref{prop:criterion}(ii)
gives $R_q(\delta;r)\leq H_q(r)+\phi(G)$. Taking the infimum over $G$ (or,
equivalently, over $W$) proves the bound.

(a) The first identity is Lemma~\ref{lem:gram} in both directions. For the
second, $\Tr G=1$ gives
$D_q(G\,\|\,\bigoplus_xG_{xx})=-S_q(G)+\sum_xS_q(G_{xx})=-\phi(G)$, and
$\sum_{x,y}\|(G^{1/2})_{xy}\|_F^2=\Tr G=1$ gives
$\pe(G)=\sum_{x\ne y}\|(G^{1/2})_{xy}\|_F^2$.

(b) A flag family is a single channel by Lemma~\ref{lem:flags}, and a single
channel is a family with one member. Let us reduce the number of members, while $|I| > q+1$. 
Discard zero-weight components initially and after each step.
The vectors $(v_i,\pe(v_i,W_i))\in\mathbb R^{q+1}$ are linearly dependent; 
a dependence coefficient vector $z$ satisfies $\sum_iz_i=0$, because the coordinates
of each $v_i$ sum to one. Choose the sign of $z$ so that
$\sum_iz_ib_i \leq 0$, where $b_i=\chi_q(v_i,W_i)-H_q(v_i)$, and put
\[
\tau_*=\min_{z_i<0}\frac{w_i}{-z_i}, \qquad w_i'=w_i+\tau_*z_i,
\]
which makes at least one of the weights zero.
The composition and the average error are preserved and the objective does not increase.

(c) If $G_1,G_2$ are feasible at $(r_1,\delta_1)$ and $(r_2,\delta_2)$, then
$\lambda G_1\oplus(1-\lambda)G_2$ is feasible at the convex combination with
the averaged objective (Lemma~\ref{lem:flags}); this gives joint convexity of $\UC(\delta;r)-H_q(r)$,
equivalently the asserted joint concavity of $H_q(r)-\UC(\delta;r)$.
Monotonicity in $\delta$ is clear, and permutation invariance is
Lemma~\ref{lem:sym}. Further, $\UC\ge0$ because $\chi_q\ge0$; $\UC\leq H_q(r)$ by the
channel with orthogonal output states ($\pe=0$, $\phi=0$). For
$\delta>\delta^\mathrm{Pl}_q(r)$ a constant channel ($\rho_x=\rho$ for all $x$) has
$\pe=\delta^\mathrm{Pl}_q(r)$ and $\chi_q=0$. Let $\phi(r,t)$ be the infimum of
$\phi(G)$ over $\comp(G)=r$, $\pe(G)\leq t$. It is convex, finite and
nonincreasing in $t\in[0,1]$, hence continuous on $(0,1)$, so
$\inf_{t<\delta}\phi(r,t)=\phi(r,\delta)$ for $0<\delta<1$. This
proves the claim about $\leq \delta$ and, since
\[
 \delta_q^\mathrm{Pl}(r)=1-\sum_xr_x^2 \leq 1-\frac{1}{q} = \theta_q < 1,
\]
also $\UC(\delta^\mathrm{Pl}_q(r);r)=0$. \qed

\subsection{Proofs of Corollaries~\ref{cor:unrestricted} and~\ref{cor:cw}}
\label{sec:proof-cor}

\begin{proof}[Proof of Corollary~\ref{cor:unrestricted}]
Trivially $U^{\mathrm{UC}}_q(\delta)\ge1+\inf\phi$ for the infimum taken over Gram data $G$ with $\pe(G) < \delta$. Conversely, let $G$ be any Gram datum with $\pe(G)<\delta$. Lemma~\ref{lem:sym} with the cyclic group
$\mathbb Z_q$ gives $G^{\mathbb Z_q}$ with composition rate $u_q$ and the same
$\pe,\phi$, realized by a $\mathbb Z_q$-covariant channel $\tilde W$ with
$\pe(u_q,\tilde W)=\pe(G)$ and $\chi_q(u_q,\tilde W)=1+\phi(G)$. This proves the
identities, including the one with covariant channels. For the rate bound,
let $r$ be any $n$-type. Lemma~\ref{lem:concave} gives
$\pe(r,\tilde W)\leq \pe(G)<\delta$ and $\chi_q(r,\tilde W)\leq 1+\phi(G)$.
Proposition~\ref{prop:criterion}(i), with the margin
$\eps=1-\pe(G)/\delta$ that does not depend on $r$, together with
$A_q(n,d)\leq (n+1)^q\max_r A_q(n,d;nr)$, gives
$\frac1n\log_qA_q(n,\lceil\delta n\rceil)\leq 1+\phi(G)+O(n^{-1/2})$.
\end{proof}

\begin{proof}[Proof of Corollary~\ref{cor:cw}]
The identity follows as above with the group $\Perm_{q-1}$ of permutations
fixing $0$, which maps any $G$ with $\Tr G_{00}=1-\omega$ to one with
composition rate $r_\omega$. A code of weight $w_n$ splits into at most
$(n+1)^q$ constant-composition subcodes whose composition rates $r$ satisfy
$r_0=1-\omega_n$, $\omega_n=w_n/n$. For the $\Perm_{q-1}$-covariant channel
$\tilde W$, Lemma~\ref{lem:concave} gives
$\pe(r,\tilde W)\leq \pe(r_{\omega_n},\tilde W)$ and
$\chi_q(r,\tilde W)\leq \chi_q(r_{\omega_n},\tilde W)$. As $\omega_n\to\omega$,
continuity gives $\pe(r_{\omega_n},\tilde W)\to\pe(G)<\delta$ and
$\chi_q(r_{\omega_n},\tilde W)\to h_q(\omega)+\phi(G)$.
For all sufficiently large $n$,
\[
 \pe(r_{\omega_n},\widetilde W)
 \leq \frac{\pe(G)+\delta}{2}.
\]
Analogously to the previous proof, Proposition~\ref{prop:criterion}(i) gives the desired bound.
\end{proof}

\section{Strict improvement between consecutive levels}
\label{sec:strict}

For an integer $m\geq 1$ and $\delta>0$, define $B_q^{(m)}(\delta;r)$ as the infimum
in Theorem~\ref{thm:main}(b) restricted to channels
$W_a:x\mapsto\rho_{a,x}$ with $\rank \rho_{a,x}\leq m$ for
all $a,x$. We will call it the \emph{flagged Horn-type bound at level $m$}.
By Theorem~\ref{thm:main}(b) we may use $q+1$ flag labels, allowing zero weights: the reduction in
Theorem~\ref{thm:main}(b) only changes weights and discards channels,
so it preserves the rank restriction.
We also set $B_q^{(m)}(0;r)=H_q(r)$; indeed, \eqref{eq:fano} and the orthogonal-output channel give
\[
 H_q(r)-h_q(\delta)
 \le B_q^{(m)}(\delta;r)\le H_q(r)
 \qquad(0<\delta\le\theta_q).
\]

Gay, Jeronimo and Liu~\cite[Proposition~8.6]{GJL26} prove a related
strictness result for their binary Horn--channel hierarchy:
if the closed optimization at one level has a minimizing pair
$(K,L)$ with both matrices positive definite, then the next level
gives a strictly smaller bound. Their proof uses the entropy gain
of order $t\log(1/t)$ produced by opening an additional spectral
direction.

\begin{theorem}[Strict improvement between consecutive levels]
\label{thm:strict}
For every $q\geq 2$, integer $m\geq 1$, composition $r\in\Delta_{q-1}$, and
$0<\delta<\delta_q^\mathrm{Pl}(r)$,
\[
 B_q^{(m+1)}(\delta;r)<B_q^{(m)}(\delta;r).
\]
In particular, $\UC(\delta;r)<B_q^{(m)}(\delta;r)$ at every finite level $m$.
\end{theorem}

\begin{proof}
Throughout the proof, the rank constraint is imposed on
each component output $\rho_{a,x}$, not on the combined
flagged output. Every additional channel is introduced
as a separate flag component.

We will improve an optimal level-$m$ feasible flag family by increasing the rank
of one output state. Whenever we add a flag, we use
Theorem~\ref{thm:main}(b) to reduce to $q+1$ flags, preserving composition,
error, and the rank bound without increasing the objective.

\emph{Choosing an optimizer.}
At each fixed level, Lemma~\ref{lem:gram} represents each component by
a normalized Gram matrix on $q$ slots of size $m$, filling smaller slots
with zeros. The $q+1$ matrices and their weights range over a compact set,
and composition, error, and the objective are continuous.
Thus the constraint $\leq \delta$ gives an attained minimum; orthogonal
pure outputs ensure feasibility. Adding a channel with orthogonal pure outputs
at composition $r$ with weight $\varepsilon>0$ makes the error strictly less
than $\delta$, and the objective value tends to that minimum as $\varepsilon\to0$.
Hence the closed minimum equals the original strict infimum at every level.

Choose a level-$m$ minimizer with objective value
$\alpha=B_q^{(m)}(\delta;r)$ and average PGM symbol error
\[
 E:=\sum_a w_a\pe(v_a,W_a)\le\delta.
\]
By Lemma~\ref{lem:flags}, $\alpha$ is the Holevo
information of its flagged channel. 
If $\alpha=0$, all output states of its flagged channel
corresponding to inputs $x$ with $r_x>0$ coincide,
giving $E=\delta_q^\mathrm{Pl}(r)>\delta$, a contradiction;
thus $\alpha>0$.
The constant pure-output channel at composition $r$ has objective value zero and error
$\delta_q^\mathrm{Pl}(r)$. If $E<\delta$, adding a small weight of this channel
would decrease the objective value while keeping the error below $\delta$.
Therefore $E=\delta$.

\emph{The selected Gram matrix is nonsingular.}
Choose a component with positive weight $w$ and positive error, which
exists since $E=\delta>0$, and denote its composition by $v$ and outputs by
$\rho_x$. Omit inputs $x$ with $v_x=0$, to which pure states may be assigned.
Let $G$ be the Gram datum supplied by Lemma~\ref{lem:gram}
with slot dimensions $d_x:=\rank\rho_x\le m$; then each diagonal block $G_{xx}$ is positive definite.
We show that $G\succ0$: otherwise a perturbation within the same level
saves enough error to lower the objective value by adding the constant channel.

Suppose $P$, the orthogonal projection onto $\ker G$, is nonzero.
Set $\widehat G=\bigoplus_xG_{xx}\succeq\lambda I$ for some $\lambda>0$
and $G_t=(1-t)G+t\widehat G$, where $0<t<1$.
Since $(G_t)_{xx}=G_{xx}$, Lemma~\ref{lem:gram}
realizes $G_t$ by states of ranks
$\rank (G_t)_{xx}=d_x\leq m$.
Thus this perturbation stays at level $m$, although
the rank of the full Gram matrix increases.
The composition and diagonal blocks are unchanged. Operator monotonicity
of the square root, followed by comparison on each eigenspace of $G$, gives
\[
 G_t^{1/2}\succeq\bigl((1-t)G+t\lambda I\bigr)^{1/2}
 \succeq\sqrt{1-t}\,G^{1/2}+\sqrt{\lambda t}\,P.
\]
Since $\Tr G=1$, we have $G^{1/2}\succeq G$, so
$(G^{1/2})_{xx}\succeq G_{xx}\succ0$. Some $P_{xx}$ is nonzero because
$P\ne0$; hence $b:=\sum_x\Tr((G^{1/2})_{xx}P_{xx})>0$.
Compressing to each slot and using $\Tr A^2\geq \Tr B^2$ for
$A\succeq B\succeq0$, we obtain
\[
 \pe(G_t)\leq \pe(G)+t\bigl(1-\pe(G)\bigr)
 -2b\sqrt{\lambda t(1-t)}.
\]
The diagonal blocks are fixed, so
$\phi(G_t)-\phi(G)=S_q(G_t)-S_q(G)$.
Applying the entropy bounds for a mixture~\cite[Eq.~(11.86), Theorem~11.10]{NC10}
to the states $G,\widehat G$ with weights $1-t,t$ gives
\[
 t\bigl(S_q(\widehat G)-S_q(G)\bigr)
 \leq S_q(G_t)-S_q(G)
 \leq t\bigl(S_q(\widehat G)-S_q(G)\bigr)+\beta_q(t),
\]
where $\beta_q(t)=-t\log_qt-(1-t)\log_q(1-t)$.
Hence $\phi(G_t)-\phi(G)=O(t\log(1/t))$.

Replacing this component therefore gives average PGM error $E_t<\delta$
and objective value $\alpha_t=\alpha+O(t\log(1/t))$, with $\delta-E_t$ at least a positive
multiple of $\sqrt t$. Add the constant channel at composition $r$ with weight
\[
 s_t=\frac{\delta-E_t}{\delta_q^\mathrm{Pl}(r)-E_t}.
\]
The error returns to $\delta$, while the objective value becomes
\[
 (1-s_t)\alpha_t=\alpha-\alpha s_t+O\bigl(t\log(1/t)\bigr)<\alpha
\]
for small $t$, since $s_t$ is at least a positive multiple of $\sqrt t$.
All component ranks remain at most $m$, contradicting minimality.
Thus $G\succ0$.

\emph{Increasing one output rank.}
Realize $G$ on its slot space using $R=G^{1/2}$ as in Lemma~\ref{lem:gram}. 
Denote the resulting output states again by $\rho_x$; their average output state is $\sum_xv_x\rho_x=G\succ0$.
There are at least two inputs $x$ with $v_x>0$, since the error is positive,
so each output state has rank smaller than the output dimension.
We can therefore increase one output rank within this space,
keeping the average state nonsingular.
Choose an input $j$ with $v_j>0$ and a unit vector
$z\perp \supp \rho_j$, and replace only that state by
\[
 \rho_{j,t}=(1-t)\rho_j+tzz^*,\qquad 0<t<1.
\]
Its rank increases by one and its entropy by $t\log_q(1/t)+O(t)$.
The perturbed average output state is
\[
 \rho_{v,t}=G+v_jt(zz^*-\rho_j).
\]
It remains positive definite for sufficiently small $t$.
Thus $S_q(\rho_{v,t})$ and $\rho_{v,t}^{-1/2}$ vary smoothly,
and the PGM error changes by $O(t)$.
The total objective value and
error consequently satisfy
\[
 \alpha_t=\alpha-wv_jt\log_q(1/t)+O(t),\qquad E_t=\delta+O(t).
\]
If $E_t>\delta$, add a channel with orthogonal pure outputs at composition $r$ with
weight $1-\delta/E_t=O(t)$. This restores error $\delta$ at a cost
of $O(t)$ in the objective; if $E_t\leq \delta$, no correction is needed.
The objective value remains strictly below $\alpha$ for small $t$, and all component output ranks are at most $m+1$. The equality of the closed and strict objective values
proved above now yields
$B_q^{(m+1)}(\delta;r)<\alpha=B_q^{(m)}(\delta;r)$.
Finally, $\UC\leq B_q^{(m+1)}$ gives the last assertion.
\end{proof}

\section{Transport and shortening}
\label{sec:closure}

\subsection{Closure operators}

For $r,s\in\Delta_{q-1}$, the permutation-transport cost
from $r$ to $s$~\cite{ABCR26} is
\[
 K_q(r,s)
 =\min_{P_{\Pi\mid X}:\,\Pi(X)\sim s} I_q(X;\Pi),
 \qquad X\sim r,
\]
where $\Pi$ is a random permutation of $\A$, allowed to depend on $X$,
and $I_q$ denotes mutual information in base-$q$ units.
The permutation-transport closure of a bound $B$ is
\[
 (\Pp B)(\delta;r)
 =\inf_{s\in\Delta_{q-1}}\{K_q(r,s)+B(\delta;s)\}.
\]
The Laihonen--Litsyn shortening inequality~\cite[Theorem~1]{LL98} yields, for an
unrestricted bound $U:[0,1]\to\mathbb R$, the generalized-shortening
closure used in~\cite[\S5.6]{ABCR26}:
\begin{equation}\label{eq:shortening}
 (\Ss U)(\delta)
 =\inf \left\{\alpha\left[1-h_q\!\left(\min\left\{t,\frac{q-1}{q}\right\}\right)\right]
 +(1-\alpha)U(\gamma)\right\},
\end{equation}
where the infimum is taken over triples $(\alpha,\gamma,t)\in[0,1]^3$ such that $2\alpha t+(1-\alpha)\gamma=\delta$.

\subsection{Closure at each level}
In this subsection we show that every level $B_q^{(m)}$, as well as the limit bound $\UC$, are fixed points of the operators $\Pp$ and $\Ss$. Explicit characterization of the fixed points gives theoretical corollaries on our bounds.

\begin{theorem}\label{thm:horn-closure}
For every integer $m\geq 1$,
\[
 \Pp B_q^{(m)}=B_q^{(m)},\qquad
 \Ss[B_q^{(m)}(\cdot;u_q)]=B_q^{(m)}(\cdot;u_q).
\]
\end{theorem}

\begin{proof}
\emph{Transport.}
Fix $m\geq 1$, $r,s\in\Delta_{q-1}$, $\delta>0$ and $\varepsilon>0$. Choose a feasible flag family for $B_q^{(m)}(\delta;s)$
 whose objective value is within $\varepsilon$ of the infimum, and let $V$ be its flagged channel from
 Lemma~\ref{lem:flags}.
The lemma identifies their objective with $\chi_q(s,V)$, so
\[
 \pe(s,V)<\delta,\qquad
 \chi_q(s,V)<B_q^{(m)}(\delta;s)+\varepsilon.
\]
Choose an optimal joint law $(X,\Pi)$ in the definition of $K_q(r,s)$:
$X\sim r$, $Y=\Pi(X)\sim s$, and $I_q(X;\Pi)=K_q(r,s)$.
The new channel $\widetilde V$ chooses $\Pi$ conditionally on its input
$X=x$, sends $\Pi(x)$ through $V$, and reveals $\Pi$ together with the
original flag of $V$.
Write $\mu_\pi=\Pr\{\Pi=\pi\}$ and let $s_\pi$ be the conditional
law of $Y$ given $\Pi=\pi$, omitting zero weights. Then
$\sum_\pi\mu_\pi s_\pi=s$, and each $s_\pi$ is supported on
$\supp s$.
By Lemmas~\ref{lem:flags}, \ref{lem:sym}, and~\ref{lem:concave},
\begin{align*}
 \pe(r,\widetilde V)
 &=\sum_\pi\mu_\pi\pe(s_\pi,V)
 \leq \pe(s,V)<\delta,\\
 \chi_q(r,\widetilde V)
 &=I_q(X;\Pi)+\sum_\pi\mu_\pi\chi_q(s_\pi,V)
 \leq K_q(r,s)+\chi_q(s,V).
\end{align*}
To check the level constraint, denote the original component states by
$\rho_{a,y}$. The channel $V$ keeps its conditional flag probabilities
and these states when used at $s_\pi$; only the composition changes.
Conditional on the old flag $a$ and the new flag $\pi$, the output state
is $\rho_{a,\pi(x)}$, of rank at most $m$.
Inputs of zero conditional probability may be assigned pure states.
Reducing to at most $q+1$ flags as in Theorem~\ref{thm:main}(b)
preserves the composition, error, and component ranks, and cannot
increase the objective. Consequently,
\[
 B_q^{(m)}(\delta;r)
 \leq \chi_q(r,\widetilde V)
 <K_q(r,s)+B_q^{(m)}(\delta;s)+\varepsilon.
\]
The limit pass gives
\[
 B_q^{(m)}(\delta;r)
 \leq K_q(r,s)+B_q^{(m)}(\delta;s).
\]
Taking the infimum over $s$ gives $B_q^{(m)}\leq \Pp B_q^{(m)}$;
the choice $s=r$ and $K_q(r,r)=0$ give the converse inequality.

At $\delta=0$, transport stability follows from $K_q(r,s)\ge H_q(r)-H_q(s)$ and the choice $s=r$.

\emph{Shortening.}
Put $U(f)=B_q^{(m)}(f;u_q)$ for $f\geq 0$.
Mixing two feasible flagged channels at composition $u_q$
and retaining the choice as a flag averages both their errors and
their information.
The component ranks remain at most $m$, and the number of flags can
again be reduced to $q+1$. Applying this construction to nearly optimal
channels at two error thresholds proves that $U$ is convex.
At distance zero use orthogonal pure output states, with error zero and information $1=U(0)$. 
For component thresholds $f_1,f_2\geq0$ and weights $\lambda,1-\lambda$, the resulting threshold is
$f=\lambda f_1+(1-\lambda)f_2$. If $f>0$, at least one positively weighted component has
a positive threshold, so the average error is strictly below $f$.

For $t\in[0,1]$, put $f_t=\min\{2t,1\}$, $\tau=\min\{t,(q-1)/q\}$ and consider the $q$-ary classical
channel that keeps the input symbol with probability $1-\tau$ and
otherwise replaces it uniformly by one of the other $q-1$ symbols.
Its output symbol $Z$ is a visible flag, with a one-dimensional
state inside each flag, so this channel is feasible already at level~$1$. Explicitly, in Lemma~\ref{lem:flags} take
\[
w_z=\frac1q,\qquad
v_x^{(z)}=
\begin{cases}
1-\tau,&x=z,\\
\tau/(q-1),&x\ne z,
\end{cases}
\qquad W_z:x\mapsto[1].
\]
Then $\sum_z w_zv^{(z)}=u_q$, and every component output
state has rank one.
At uniform input its Holevo information and PGM error are
\[
 1-h_q(\tau),\qquad
 1-\left((1-\tau)^2+\frac{\tau^2}{q-1}\right)
 =2\tau-\frac{q}{q-1}\tau^2.
\]
For $t>0$, the displayed error is strictly less than $2t$
and at most $(q-1)/q<1$, hence strictly less than $f_t$.
At $t=0$, use $U(0)=1=1-h_q(0)$. Thus
\[
 U(f_t) \leq 1-h_q(\tau)\qquad(0\leq t\leq 1).
\]
For any parameters in~\eqref{eq:shortening}, monotonicity and convexity now give
\[
   \alpha f_t+(1-\alpha)\gamma
   \le 2\alpha t+(1-\alpha)\gamma=\delta
\]
and
\[
U(\delta) \leq U\bigl(\alpha f_t+(1-\alpha)\gamma\bigr) \leq \alpha U(f_t)+(1-\alpha)U(\gamma) \leq \alpha (1-h_q(\tau)) + (1-\alpha)U(\gamma).
\]
Taking the infimum over these parameters gives $U\leq\Ss U$.
The choice $\alpha=0$, $\gamma=\delta$ gives the converse
inequality for every $0\leq\delta\leq1$.
\end{proof}

Every finite family of channels has bounded component ranks, so
\begin{equation}\label{eq:horn-universal-infimum}
 \UC(\delta;r)=\inf_{m\geq 1}B_q^{(m)}(\delta;r),
 \qquad\delta\geq 0.
\end{equation}

\begin{corollary} \label{cor:universal}
One has
\[
 \Pp\UC=\UC, \qquad \Ss U_q^{\mathrm{UC}}=U_q^{\mathrm{UC}}.
\]
\end{corollary}

We say that a nonincreasing probability vector $y$ \textit{majorizes} nonincreasing probability vector $x$, and write $x \preceq y$ when
\[
 \sum_{i=1}^j x_i \leq \sum_{i=1}^j y_i 
 \quad(1\leq j < k),
\]
where $k$ is the common length of $x$ and $y$. A permutation-invariant function $f$ on probability vectors of length $k$ is \textit{Schur-concave} if $f(x) \geq f(y)$ provided that $x\preceq y$.

\begin{corollary} \label{cor:SchurConcavity}
For every $m \geq 1$ and $\delta>0$ the functions $B_q^{(m)}(\delta;r)$ and $H_q(r) - B_q^{(m)}(\delta;r)$
 are Schur-concave in $r$. The same holds for $\UC(\delta;r)$ and $H_q(r) - \UC(\delta;r)$.
\end{corollary}

\begin{proof}
By~\cite[Corollary 3.3]{ABCR26}, for $s\preceq r$ one has $K_q(r,s)=0$ and $K_q(s,r)=H_q(s)-H_q(r)$.
This immediately implies
\[
0\leq B_q^{(m)}(\delta;s)-B_q^{(m)}(\delta;r)
\leq H_q(s)-H_q(r)
\]
and the same inequality for $\UC$.
\end{proof}

\begin{corollary}[Consequences of shortening stability]
\label{cor:shortening-shape}
Let $q\geq 2$, put $\theta_q=(q-1)/q$, and let
$U(\delta)=B_q^{(m)}(\delta;u_q)$ for any integer $m\geq 1$, or
$U(\delta)=U_q^{\mathrm{UC}}(\delta)$.
Then the functions
\[
 \delta\longmapsto\frac{1-U(\delta)}{\delta},
 \qquad
 \delta\longmapsto\frac{U(\delta)}{2\theta_q-\delta}
\]
are nonincreasing on $(0,1)$. Equivalently, for $0<d<\delta<1$,
\begin{equation}\label{eq:shortening-two-point}
 1-\frac{\delta}{d}\bigl(1-U(d)\bigr)
 \leq U(\delta)
 \le\frac{2\theta_q-\delta}{2\theta_q-d}\,U(d).
\end{equation}
Moreover, at every $0<\delta<1$ and for every
$p\in[U'_-(\delta),U'_+(\delta)]$,
\[
 p\le0,\qquad U(\delta)-\delta p \leq 1-\log_q\!\left(1+(q-1)q^{2p}\right).
\]
Here $U'_-$ and $U'_+$ denote the left and right derivatives.
The same conclusions hold for $U(\delta)=F_q(\delta;u_q)$,
defined in~\eqref{eq:flagged-qms}.
\end{corollary}

\begin{proof}
Theorem~\ref{thm:horn-closure} and Corollary~\ref{cor:universal}
give $\Ss U=U$; for $F_q$, use
Proposition~\ref{prop:compact-closure}.
Apply the secant characterization in
Theorem~\ref{thm:shortening-fixed-points} with $z=0$ and
$z=2\theta_q$, for which $c_q(z)=1$ and $0$, respectively.
This gives the two monotonicity assertions and
\eqref{eq:shortening-two-point}.
These channel bounds are convex by mixing flag families,
nonincreasing, and continuous on $[0,1]$; continuity at zero follows
from the Fano bound, and they vanish on $[\theta_q,1]$.
The slope assertion follows from
Theorem~\ref{thm:shortening-fixed-points} and the convex-case
observation following its proof.
\end{proof}

\section{Special cases}
\label{sec:special}

We compare the channel optimization of Theorem~\ref{thm:main}
with the bounds obtained by Ben-Haim and Litsyn~\cite{BHL07},
OpenAI~\cite{OAI26}, Alrabiah and Guruswami~\cite{AG26},
Gay, Jeronimo and Liu~\cite{GJL26}, and
Salmon~\cite{Sal26,salmon2026rate}, using the formulations
collected in ABCR~\cite[\S6]{ABCR26}.

For the explicit channel and moving-subspace bounds from~\cite{OAI26,AG26,GJL26}, our
comparisons are established at the level of the optimization
domains: we embed the relevant feasible parameter families into
our channel and flag families, with the corresponding objective
values and error constraints preserved or improved.
Where necessary, these realizations use limits or the transport
and shortening constructions of Section~\ref{sec:closure}.
The resulting inequalities between optimized bounds are
consequences of these explicit realizations.
The provided comparisons and existing relations between the bounds are illustrated in Fig.~\ref{fig:channel-comparisons}.
The finite matrix parametrizations of our families are given in
Sections~\ref{sec:recipe-qms} and~\ref{sec:recipe-horn}.

\subsection{Fully permutation-covariant \texorpdfstring{$q$}{q}-ary channels}
\label{sec:qmqc}

We consider the family of $\Perm_q$-covariant (fully permutation-covariant, FPC-$q$)
$q$-ary channels on $\mathbb C^q$.
This family contains the $q$-ary mixed channels ($q$-MC)
of Alrabiah and Guruswami~\cite[\S7, (55)]{AG26}.
For $q=2$, up to a common output unitary, it coincides
with the mixed-qubit channel family from \cite[\S5.1, Remark~5.2]{AG26}.

In the definitions of FPC-$q$ and qMS below, covariance
is always with respect to the natural permutation
representation $P_\sigma e_x=e_{\sigma(x)}$ on $\mathbb C^q$.

Let $e_0,\ldots,e_{q-1}$ be the standard basis of $\mathbb C^q$. Set
$f_x=(q-1)^{-1/2}\sum_{y\ne x}e_y$ and
$\Pi_x=I-e_xe_x^*-f_xf_x^*$. Given
\[
 B=\begin{pmatrix}a&z\\\overline z&b\end{pmatrix}\succeq0,
 \qquad \ell\geq 0,\qquad a+b+(q-2)\ell=1,
\]
define the \emph{fully permutation-covariant $q$-ary channel} (FPC-$q$) by
\begin{equation}\label{eq:qmqc-seed}
 \rho_x=a e_xe_x^*+z e_xf_x^*+\overline z f_xe_x^*
 +b f_xf_x^*+\ell\Pi_x.
\end{equation}
Thus positivity means $a,b$ are nonnegative reals and $|z|^2\leq ab$. The parameter $z$ may be complex; $a,b,\ell$ are real. For $q=2$, $\Pi_x=0$ and the
scalar $\ell$ is omitted; the trace constraint is $a+b=1$.

\begin{proposition}[Covariance characterization]\label{prop:qmqc}
The FPC-$q$ channels are exactly the channels covariant
under the natural permutation representation of $\Perm_q$
on $\mathbb C^q$. They contain the $q$-MC family of Alrabiah and Guruswami
\cite[\S7, (55) and Table~3]{AG26}.
\end{proposition}

\begin{proof}
The state $\rho_0$ must commute with all permutations fixing $0$.
For $q\geq 3$ their representation decomposes into the two trivial
summands spanned by $e_0,f_0$ and the standard representation on their
orthogonal complement. The latter is irreducible and inequivalent to the
trivial representation, also when $q=3$, when it is the one-dimensional
sign representation. The commutant therefore consists of an arbitrary
complex $2\times2$ block and a scalar on the complement. Positivity and
trace one give exactly \eqref{eq:qmqc-seed}. Permuting $\rho_0$ generates
all outputs, independently of the chosen permutation. For $q=2$ the
stabilizer is trivial, so the state $\rho_0$ is an arbitrary qubit state
and $\rho_1$ is its conjugate by the transposition.

For the explicit inclusion, use the pure-state and classical-preprocessing parameters in \cite[\S7, (55)]{AG26}, written here as
$0 \leq t, \eta \leq \theta_q=(q-1)/q$, and put
\[
 a_t=\sqrt{1-t},\quad b_t=\sqrt{t/(q-1)},\quad
 \varphi=\binom{a_t}{\sqrt t},\quad
 w=\binom{b_t}{(a_t+(q-2)b_t)/\sqrt{q-1}}.
\]
The substitution is
\begin{equation}\label{eq:mc-embedding}
 B=(1-\eta)\varphi\varphi^*+\eta ww^*,\qquad
 \ell=\frac{\eta}{q-1}(a_t-b_t)^2.
\end{equation}
Indeed, AG's vectors are $\psi_y=(a_t-b_t)e_y+b_t\1$ and its outputs are
$(1-\eta)\psi_x\psi_x^*+\eta(q-1)^{-1}\sum_{y\ne x}\psi_y\psi_y^*$.
The vector $\psi_x$ has coordinates $\varphi$ on
$\langle e_x,f_x\rangle$; each remaining vector projects there as $w$.
Their complementary projections sum to zero and their outer products sum
to $(a_t-b_t)^2\Pi_x$. This gives \eqref{eq:mc-embedding}, including its
normalization, directly. The same calculation applies at $q=2$ with no
scalar sector.
\end{proof}

Rank-one $B$ with $\ell=0$ gives the covariant pure-state specialization.
In \eqref{eq:mc-embedding}, $\eta=0$ recovers AG's pure-state channel and $t=0$
its classical symmetric channel.

\subsection{General \texorpdfstring{$q$}{q}-ary moving-subspace channels}
\label{sec:qms}

Let $\mathcal W_q$ consist of the channels $x\mapsto\rho_x$ on
$\mathbb C^q$ covariant under the stabilizer $\Stab(0)$ of the symbol $0$ in $\Perm_q$.
We call this the $q$-ary moving-subspace (qMS) family. 
By definition $\text{FPC-}q \subseteq\mathcal W_q$,
and $\mathcal W_2$ contains every binary qubit channel.

\begin{proposition}[Parametrization of qMS channels]\label{prop:qms-seeds}
For $q\geq 3$, set $E_0=[e_0\ f_0]$ and
$\Pi_0=I-E_0E_0^*$. Every channel in $\mathcal W_q$ has a distinguished state 
\[
 \rho_0=E_0 B_0E_0^*+\ell_0\Pi_0,\qquad
 B_0\in\mathbb C^{2\times2},\quad B_0\succeq0,\quad
 \ell_0\geq 0,\quad \Tr B_0+(q-2)\ell_0=1.
\]
For $q\geq 4$, put $g_1=(q-2)^{-1/2}\sum_{j=2}^{q-1}e_j$,
$E_1=[e_0\ e_1\ g_1]$, and $\Pi_1=I-E_1E_1^*$. The output state $\rho_1$ has the form
\[
 \rho_1=E_1 B_1E_1^*+\ell_1\Pi_1,\qquad
 B_1\in\mathbb C^{3\times3},\quad B_1\succeq0,\quad
 \ell_1\geq 0,\quad \Tr B_1+(q-3)\ell_1=1.
\]
For $q=3$ use the same $E_1=[e_0\ e_1\ e_2]$ and an arbitrary
$3\times3$ state $B_1$; the scalar sector is absent. For $q=2$,
$\rho_0,\rho_1$ are two independent arbitrary $2\times2$ states.
For $x\ne0$, choose $\sigma_x\in\Stab(0)$ with $\sigma_x(1)=x$ and set
$\rho_x=P_{\sigma_x}\rho_1P_{\sigma_x}^*$.
These choices give exactly $\mathcal W_q$, with no further compatibility
conditions between $\rho_0$ and $\rho_1$. 
\end{proposition}
\begin{proof}
The distinguished state $\rho_0$ commutes with the whole subgroup $\Perm_{q-1}$,
so its commutant is the one used in Proposition~\ref{prop:qmqc}.
The nondistinguished state $\rho_1$ needs to commute only with the stabilizer of both
$0$ and $1$. For $q\geq 4$ this is $\Perm_{q-2}$, with three trivial summands
$e_0,e_1,g_1$ and one standard summand of dimension $q-3$.
By Schur's lemma~\cite[Proposition~2.3.9 and Corollary~2.3.10]{EGH11}, an operator commuting with this representation has no blocks between the trivial and standard summands and acts as a scalar on the standard summand.
Its restriction to the three-dimensional fixed subspace
$\spann\{e_0,e_1,g_1\}$ is arbitrary.
This gives the block $B_1$ and scalar $\ell_1$ in the stated representation. At $q=4$ the
standard summand is a nontrivial sign representation, so the formula still
holds. At $q=3$ this stabilizer is trivial and the whole state is arbitrary.
Two permutations sending $1$ to $x$ differ by a stabilizer element;
therefore they give the same $\rho_x$. Conversely, states satisfying these stabilizer commutation conditions generate
a $\Stab(0)$-covariant channel, since the same stabilizer
argument verifies every covariance identity. At $q=2$ the subgroup fixing
$0$ is trivial, imposing no condition on either output. 
\end{proof}

\subsection{Flagged qMS, compactness and closure}
\label{sec:flagged-qms}

The same qMS states support a single optimization for every composition $r$.
Allow a component channel $W_a\circ\pi_a$, where $W_a\in\mathcal W_q$,
$\pi_a\in\Perm_q$, and its composition $v_a$ is arbitrary.
Write $\Phi_a=\chi_q(v_a,W_a\circ\pi_a)-H_q(v_a)$. Define \textit{the flagged qMS bound} as
\begin{equation}\label{eq:flagged-qms}
 F_q(\delta;r) := H_q(r)+\min\left\{\sum_aw_a\Phi_a:
 \begin{array}{l}
 w_a\geq 0,\quad\sum_aw_a=1,\\
 \sum_aw_av_a=r,\quad\sum_aw_a\pe(v_a,W_a\circ\pi_a)\leq \delta
 \end{array}\right\}.
\end{equation}
The weights are the flag probabilities; the last two constraints specify
average composition and average error. The component-reduction argument
of Theorem~\ref{thm:main}(b) applies unchanged and leaves at most $q+1$
actual components. The qMS states, compositions and weights then range over
compact sets, and entropy and PGM error are continuous, also at singular
boundaries (use their Gram expressions). Thus the minimum in~\eqref{eq:flagged-qms} is
attained. For $\delta>0$ its value equals the infimum with strict error:
mix a closed-feasible flag family with an arbitrarily small weight of
the orthogonal-output channel at the same composition $r$, then reduce the
components again. The error becomes strictly smaller unless it was
already zero, and the objective converges to its original value. Every component acts on $\mathbb C^q$, so its component output ranks
are at most $q$. Consequently,
\[
 R_q(\delta;r)\leq \UC(\delta;r) \leq B_q^{(q)}(\delta;r)\leq F_q(\delta;r).
\]

\begin{proposition}[Strict improvement over flagged qMS]
\label{prop:compact-strict}
For $q\geq 2$, $r\in\Delta_{q-1}$ and
$0<\delta<\delta_q^{\mathrm{Pl}}(r)$,
\[
 \UC(\delta;r)<B_q^{(q)}(\delta;r)<F_q(\delta;r).
\]
\end{proposition}
\begin{proof}
Take a closed-error minimizer for $F_q(\delta;r)$. As in the proof
of Theorem~\ref{thm:strict}, its value is positive and its average
error equals $\delta$: a constant pure-output channel belongs to
$\mathcal W_q$ and would improve a minimizer with a slack constraint.
Choose a positive-weight component with positive error, denote its
composition by $v$ and outputs by $\rho_x$, and form its minimal Gram matrix
$G$, omitting inputs $x$ with $v_x=0$. Every component output rank
is at most $q$, and $\rank G\leq q$, since the
original component acts on $\mathbb C^q$.

If $G$ is singular, the perturbation
$G_t=(1-t)G+t\bigoplus_xG_{xx}$ from the proof
of Theorem~\ref{thm:strict} preserves the
output ranks. Its error saving is at least a positive multiple
of $\sqrt t$, while its information cost is $O(t\log(1/t))$.
Adding a constant pure-output flag to restore average error $\delta$
therefore produces a strictly better level-$q$ feasible flag family.

If $G$ is nonsingular, its size is
$N=\sum_{x:v_x>0}\rank \rho_x
=\rank G\leq q$.
There are at least two inputs $x$ with $v_x>0$, so every output rank is
at most $q-1$. The rank-increasing perturbation and the orthogonal-output
correction in the proof of Theorem~\ref{thm:strict} therefore remain
within level $q$ and strictly improve the objective.
The equality of the closed and strict level-$q$ values, established
in that proof, gives $B_q^{(q)}<F_q$. The inequality
$\UC<B_q^{(q)}$ is Theorem~\ref{thm:strict}.
\end{proof}

\begin{proposition}[Closure of flagged qMS]
\label{prop:compact-closure}
For $0<\delta\leq 1$, the bound $F_q$ satisfies
\[
 \Pp F_q=F_q,\qquad \Ss[F_q(\cdot;u_q)]=F_q(\cdot;u_q).
\]
\end{proposition}
\begin{proof}
For transport, use the construction in the proof of
Theorem~\ref{thm:horn-closure} with $V$ the flagged channel
of a feasible family in~\eqref{eq:flagged-qms}.
Conditional on the new permutation $\pi$, each original
component $W_a\circ\pi_a$ becomes $W_a\circ\pi_a\circ\pi$,
with its induced conditional composition and flag weight.
Thus it remains in the allowed family.

For shortening, mix flagged channels at composition $u_q$
and use the classical symmetric channel with parameter
$\tau=\min\{t,(q-1)/q\}$ from the same proof.
This channel belongs to $\mathcal W_q$ and has information
$1-h_q(\tau)$ and PGM error $2\tau-q\tau^2/(q-1)$.
Nested flags can be combined into a single flag indexed by the pair of labels, and reduction to $q+1$
components preserves the constraints without increasing
the objective.
\end{proof}

\subsection{The selected nonbinary bounds}
\label{sec:envelope}

We now identify the six nonbinary inputs selected in ABCR
\cite[\S6]{ABCR26}, together with the binary moving-subspace
inputs. The identifications are at their stated compositions; closure is
applied once after the constructions. Constant pure-output channels are available in
$\mathcal W_q$ (take all outputs $e_0e_0^*$). Such a component has
$\chi_q=0$ and error $\delta^\mathrm{Pl}_q(v)=1-\sum_xv_x^2$ at composition $v$.

\paragraph{Nonbinary mixed channels.}
Let $U_q^{\MC}$ be the explicit $q$-MC rate bound of
\cite[\S7, Table~3]{AG26}, recorded in
\cite[\S5.6]{ABCR26}. Proposition~\ref{prop:qmqc} and
\eqref{eq:mc-embedding} give
\begin{equation}\label{eq:compare-mc}
 F_q(\delta;u_q)\leq U_q^{\MC}(\delta).
\end{equation}
Indeed, the defining channels have PGM error $\delta$ at composition $u_q$
for $0<\delta<\theta_q$ and are closed-feasible for $F_q$.
Orthogonal and constant pure outputs give the endpoint and zero-rate
cases, and limiting parameters are covered by compactness.
The specialization to the classical symmetric and pure-state channels,
and their recovery of the Elias--Bassalygo and first LP bounds, are
already given in \cite[\S7, Table~3]{AG26}; these are not new
comparisons here.

\paragraph{Constant-weight Ben-Haim--Litsyn (BHL).}
Here $q\geq 3$. We use the product-scheme bound of \cite{BHL07} in precisely
the form selected by ABCR \cite[\S5.4]{ABCR26}. Put
\[
 k_Q(t)=\frac{Q-1}{Q}-\frac{Q-2}{Q}t-
 \frac2Q\sqrt{(Q-1)t(1-t)},\qquad
 h(x,y)=\frac{x(1-x)-y(1-y)}{1+2\sqrt{y(1-y)}}.
\]
For $0<w<1$, the parameters of a single component satisfy
\begin{equation}\label{eq:bhl-domain}
 0\leq \eta\leq \frac{q-2}{q-1}w,\qquad
 0\leq \xi-\eta\leq \min\{w-\eta,1-w\}.
\end{equation}
Set $x=(w-\eta)/(1-\eta)$, $y=(\xi-\eta)/(1-\eta)$,
$b=(1-\eta)h(x,y)$, $\mu=w-b$ and
\[
 d_q(w;\eta,\xi)=2b+\mu k_{q-1}(\eta/\mu),\qquad
 L_q(\eta,\xi)=h_q(\xi)+\xi h_q(\eta/\xi)
 -(\xi+\eta)\log_q(q-1)+\eta\log_q(q-2).
\]
In $R_1(q,w,d)$ ABCR minimize $L_q(\eta,\xi)$ subject to
\eqref{eq:bhl-domain} and the additional admissibility conditions
\begin{equation}\label{eq:bhl-admissible}
 b\leq w-\frac{q-1}{q-2}\eta,\qquad d_q(w;\eta,\xi)\leq d.
\end{equation}
Zero-mass terms are interpreted by continuity.

\begin{theorem}[Pure-state realization of a BHL component]\label{thm:bhl}
For parameters satisfying \eqref{eq:bhl-domain}, let
$\Theta\in\mathbb R^{q\times q}$ be the symmetric matrix with
unit diagonal, $\Theta_{0s}=\Theta_{s0}=\zeta_0$ for $s\ne0$,
and $\Theta_{st}=\zeta_1$ for distinct nonzero $s,t$, where
\[
 \zeta_1=1-\frac{(q-1)\eta}{(q-2)w},\qquad
 \zeta_0^2w(1-w)=(1-\eta)^2[x(1-x)-y(1-y)],\qquad \zeta_0\geq 0.
\]
Then $\Theta\succeq0$, and the pure states
$\rho_x=\Theta^{1/2}e_xe_x^*\Theta^{1/2}$ 
define a qMS channel $W:x\mapsto\rho_x$ satisfying
\[
 \chi_q(r_w,W)=L_q(\eta,\xi),\qquad
 \pe(r_w,W)=d_q(w;\eta,\xi).
\]
Conversely, every real positive semidefinite matrix $\Theta$ of this form with
$0\leq \zeta_1\leq 1$ is obtained from \eqref{eq:bhl-domain}, up to the sign of $\zeta_0$.
\end{theorem}

\begin{proof}
Write $u=1-\eta$, $s=\sqrt{y(1-y)}$, $\ell=\eta/(q-2)$ and
$\Gamma=D_w^{1/2}\Theta D_w^{1/2}$, $D_w=\diag r_w$.
In this notation $\Theta$ is the Gram matrix of the unit output vectors $\psi_x=\Theta^{1/2}e_x$, since $\psi_x^*\psi_y=\Theta_{xy}$.
In particular, $\Tr\Theta=q$, whereas $\Gamma$ is the Gram
datum at composition $r_w$ and satisfies $\Tr\Gamma=1$.
On the nonzero-coordinate sum-zero subspace, $\Gamma=\ell I$; on
$\spann \{e_0,(q-1)^{-1/2}\sum_{j\ne0}e_j\}$ it equals
\[
 B=\begin{pmatrix}1-w&\zeta_0\sqrt{w(1-w)}\\
 \zeta_0\sqrt{w(1-w)}&w-\eta\end{pmatrix}.
\]
The trace and determinant of $B$ are $u$ and $u^2y(1-y)$, so its
eigenvalues are $u(1-y)=1-\xi$ and $uy=\xi-\eta$. Positivity follows,
and the entropy of the spectrum
$(1-\xi,\xi-\eta,\ell,\ldots,\ell)$ is $L_q$.
Moreover $\Theta$ commutes with every permutation fixing zero; so does
its square root, proving the qMS covariance of the displayed states.
For the error, put $\tau=\sqrt{u(1+2s)}$ and use
$B^{1/2}=(B+usI)/\tau$. Its diagonal entries give
\[
 (\Gamma^{1/2})_{00}^2=1-w-b,\qquad
 (\Gamma^{1/2})_{jj}=\frac{\sqrt{\mu-\eta}+\sqrt{(q-2)\eta}}{q-1}
 \quad(j\ne0).
\]
Indeed $(x+s)^2=(1+2s)(x-h(x,y))$, and the corresponding identity with
$x$ replaced by $1-x$ gives the first formula. Thus
\[
 1-\pe=1-w-b+
 \frac{(\sqrt{\mu-\eta}+\sqrt{(q-2)\eta})^2}{q-1},
\]
which, on expanding $k_{q-1}$, is exactly $\pe=2b+\mu k_{q-1}(\eta/\mu)$.
All formulas extend through zero eigenvalues by continuity.

For the converse, put $\eta=(q-2)w(1-\zeta_1)/(q-1)$ and form the same block
$B$. Its trace is $u=1-\eta>0$. Set
$y=\lambda_{\min}(B)/u$ and $\xi=\eta+uy$.
Then $0\leq y\leq 1/2$ and
\[
 y(1-y)=\frac{\det B}{u^2}\leq x(1-x),
 \qquad x=\frac{w-\eta}{u},
\]
so $y\leq \min\{x,1-x\}$, exactly the remaining condition in
\eqref{eq:bhl-domain}. The determinant recovers $\zeta_0^2$.
Conjugating $\Theta$ by $\diag(-1,1,\ldots,1)$ reverses the sign of $\zeta_0$
without changing the information or PGM error.
\end{proof}

The full two-component bound is the infimum of
\[
 h_q(\omega)-\nu h_q(w_1)-(1-\nu)h_q(w_2)
 +\nu L_q(\eta_1,\xi_1)+(1-\nu)L_q(\eta_2,\xi_2)
\]
over $0\leq \nu,w_1,w_2\leq 1$, $\nu w_1+(1-\nu)w_2=\omega$,
$0\leq d_i\leq \min\{2w_i,1\}$, $\nu d_1+(1-\nu)d_2=\delta$, and
parameter choices satisfying \eqref{eq:bhl-domain}--\eqref{eq:bhl-admissible} with
$(w,d)=(w_i,d_i)$. This is ABCR's parameterization with
$\phi=\nu w_1$. Choose component compositions $r_{w_i}$, weights
$\nu,1-\nu$, and the channels just constructed. Their average composition is
$r_\omega$, their average error is at most $\delta$, and their flag
objective is exactly the displayed expression. Consequently
\begin{equation}\label{eq:compare-bhl-cw}
 F_q(\delta;r_\omega)\leq B_q^{\BHL}(\delta,\omega).
\end{equation}
At $w_i=0$ or $1$, take limits of the trace-one Gram matrices $\Gamma_i$;
the qMS state family is compact, so these limits retain the realization.
Zero-weight components are omitted.

\paragraph{Unrestricted Ben-Haim--Litsyn (BHL, unr).}
The passage from the constant-weight bound to its unrestricted
version is the Bassalygo--Elias reduction used in
\cite[Lemma~2.3 and \S5.6]{ABCR26}.
Its channel realization is the cyclic symmetrization of
Lemma~\ref{lem:sym}. Expand each preceding feasible flag family over cyclic relabellings, with each of
the $q$ shifts having equal weight. Its average composition becomes
$u_q$, its error is unchanged, and its objective increases by
$1-h_q(\omega)$. Thus the very same two-component realization gives
\begin{equation}\label{eq:compare-bhl-unr}
 F_q(\delta;u_q)\leq 
 \inf_\omega\{1-h_q(\omega)+B_q^{\BHL}(\delta,\omega)\}
 =: U_q^{\BHL,\unr}(\delta).
\end{equation}
The orbit expansion may have more than two actual flags.

The comparisons with earlier nonbinary bounds should also be kept
separate from this realization. ABCR~\cite[\S5.6]{ABCR26}
record that unrestricted BHL includes the Aaltonen bound~\cite{Aaltonen1990} and prove
that it is no larger than Kaipa's bound~\cite{Kaipa2018} (the argument following~\cite[(48)]{ABCR26}).
They also record the inclusion of the first LP bound in $q$-MC.
No pointwise ordering between $U_q^{\MC}$ and
$U_q^{\BHL,\unr}$ is asserted here; both are retained in the envelope.

\paragraph{Relation-layer Bassalygo--Elias.}
We use the expression in \cite[\S5.2, (42)]{ABCR26}.
Here $q\geq 3$ and initially $0<\omega<1$. For its constant-weight parameters $\alpha,\beta\geq 0$,
$\alpha\leq 1-\omega$, $\alpha+\beta\leq \omega$, set
$\gamma=\omega-\alpha-\beta$. Use constant pure-output components: one with
weight $1-\omega$ and composition
\[
 v_0=\left(1-\frac\alpha{1-\omega},
 \frac\alpha{(1-\omega)(q-1)},\ldots,
 \frac\alpha{(1-\omega)(q-1)}\right),
\]
and, for each $s\ne0$, one with weight $\omega/(q-1)$ and composition
\[
 v_s(0)=\frac\alpha\omega,\quad v_s(s)=\frac\gamma\omega,
 \quad v_s(y)=\frac\beta{\omega(q-2)}\quad(y\notin\{0,s\}).
\]
Their mean is $r_\omega$. Directly summing the component entropies and
errors gives the two expressions
\begin{align*}
 \sum_aw_aH_q(v_a)
 &= (1-\omega)h_q\left(\frac\alpha{1-\omega}\right)
 +\omega H_q\left(\frac\alpha\omega,\frac\beta\omega,
 \frac\gamma\omega\right)+\beta\log_q(q-2)
 =:\Phi_q^{\mathrm{rel}},\\
 \sum_aw_a\delta^\mathrm{Pl}_q(v_a)
 &=2\alpha-\frac{q\alpha^2}{(q-1)(1-\omega)}
 +\omega-\frac{\alpha^2+\gamma^2+\beta^2/(q-2)}\omega
 =:\delta_q^{\mathrm{loc}}.
\end{align*}
Thus $F_q(\delta;r_\omega)\leq h_q(\omega)-\Phi_q^{\mathrm{rel}}$
whenever $\delta_q^{\mathrm{loc}}\leq \delta$. Optimizing gives
\begin{equation}\label{eq:compare-rel-be}
 F_q(\delta;r_\omega)\leq B_q^{\mathrm{rel\text{-}BE}}(\delta,\omega).
\end{equation}
At $\omega=0$, only the component with composition $e_0$ remains.
At $\omega=1$, admissibility forces $\alpha=0$; omit the component
of weight $1-\omega$ and use the remaining displayed compositions.

\paragraph{Constant-composition Plotkin and code--anticode.}
At composition $r$, the constant pure-output channel has
Holevo information zero and PGM error $\delta_q^\mathrm{Pl}(r)$,
while the orthogonal-output channel $\rho_x=e_xe_x^*$
has Holevo information $H_q(r)$ and PGM error zero.
These give the two branches of $B_q^{\mathrm{cc\text{-}Pl}}$, including
the boundary through the closed-error convention. Hence
\begin{equation}\label{eq:compare-cc-pl}
 F_q(\delta;r)\leq B_q^{\mathrm{cc\text{-}Pl}}(\delta;r).
\end{equation}
For the code--anticode input \cite[\S5.8]{ABCR26}, take
$0\leq z\leq r$ and $\lambda=\sum_xz_x$. Use constant pure-output components
with compositions $e_x$ of weights $r_x-z_x$, and one with composition $z/\lambda$
of weight $\lambda$. The mean component composition is $r$.
Let $\widetilde W$ be the resulting flagged channel.
By Lemma~\ref{lem:flags},
\[
 \chi_q(r,\widetilde W)=H_q(r)-\lambda H_q(z/\lambda),
 \qquad
 \pe(r,\widetilde W)
 =\lambda-\frac{\sum_xz_x^2}{\lambda}.
\]
For $M=\max_xz_x$, the PGM error $\pe(r,\widetilde W)$ is at most both $\lambda$ and
$\lambda-M^2/\lambda=2(\lambda-M)-(\lambda-M)^2/\lambda$.
It is therefore below $\delta$ under ABCR's admissibility condition
$\min\{\lambda,2(\lambda-M)\}<\delta$. Optimizing proves
\begin{equation}\label{eq:compare-cc-ca}
 F_q(\delta;r)\leq B_q^{\mathrm{cc\text{-}CA}}(\delta;r).
\end{equation} Omit the last component if
$\lambda=0$; if $M=\lambda$ it is a point mass of error zero.

\subsection{Binary specializations}
\label{sec:binary}

For $q=2$, the qMS covariance condition is empty, so
$\mathcal W_2$ consists of all binary qubit channels.
The identity between the whole-cube moving-subspace bound~\cite{OAI26} and MQC~\cite[\S5.1, Remark~5.2 and Corollary~23]{AG26},
\[
 \kappa_H(\delta)=R_{\mathrm{MQC}}(\delta),
\]
is already established by Barg~\cite[\S2, (3)]{Bar26} and also by
Gay, Jeronimo and Liu~\cite[Proposition~D.1]{GJL26}.

For the constant-weight construction of~\cite[Chapter~2, Theorem~3.4 and (4)--(7)]{OAI26}, we record the
explicit qubit realization of its objective and feasibility condition. 

\begin{proposition}[Binary layer coordinates]\label{prop:layer}
For the parameters $(\alpha,\beta,\gamma,u)$ feasible at distance
$\delta$ in \cite[Chapter~2, (4)--(7)]{OAI26}, there is a channel $W\in\mathcal W_2$ such that
\begin{align*}
 \chi_2(r_\alpha,W)
 &=h_2(u)-\alpha h_2(\beta/\alpha)
 -(1-\alpha)h_2(\gamma/(1-\alpha)),\\
 \pe(r_\alpha,W)
 &=2\alpha(1-\alpha)\bigl(1-\Lambda_{\alpha,\beta,\gamma}(u)\bigr)<\delta,
\end{align*}
where $\Lambda_{\alpha,\beta,\gamma}(u)$ is the expression in
\cite[Chapter~2, (6)]{OAI26}. Thus both the objective and the feasibility
condition of the layer bound are recovered exactly.
\end{proposition}
\begin{proof}
The parameter range has $0<u<\alpha<1/2$, so the
denominators below are nonzero. 
Set
\[
s=\frac{u(1-u-\alpha)+\beta(\alpha-\beta)
-\gamma(1-\alpha-\gamma)}{1-2u},
\qquad
t=\sqrt{s(\alpha-s)-\beta(\alpha-\beta)},
\]
and define
\[
S=\diag(u,1-u),\qquad A=\begin{pmatrix}s&t\\t&\alpha-s\end{pmatrix}.
\]
The cited parameter ranges make $t$ real. The trace and determinant
identities give $\mathrm{spec}\,A=\{\beta,\alpha-\beta\}$ and
$\mathrm{spec}(S-A)=\{\gamma,1-\alpha-\gamma\}$. Thus
$\rho_1=A/\alpha$ and $\rho_0=(S-A)/(1-\alpha)$
are states. Define $W:x\mapsto\rho_x$. Its average output state
at $r_\alpha=(1-\alpha,\alpha)$ is
$(1-\alpha)\rho_0+\alpha\rho_1=S$.
The spectra of $\rho_0$, $\rho_1$, and $S$ give the stated
Holevo information. Finally, substitution in
\[
 \pe(r_\alpha,W)=2\alpha-2\Tr(S^{-1/2}AS^{-1/2}A)
 =2\alpha-2\left(\frac{s^2}{u}
 +\frac{(\alpha-s)^2}{1-u}+\frac{2t^2}{\sqrt{u(1-u)}}\right)
\]
gives the stated error formula.
\end{proof}

Let $B_2^{\mathrm{MS\text{-}CW}}(\delta,\alpha)$ denote the infimum of
this objective at fixed $\alpha$ \cite[\S5.7]{ABCR26}, and let
$\kappa_{\mathrm{CW}}$ be the infimum of
$1-h_2(\alpha)+B_2^{\mathrm{MS\text{-}CW}}(\delta,\alpha)$ over feasible
$\alpha$ \cite[Chapter~2, (8) and Theorem~3.6]{OAI26}. For $0<\delta<1/2$ and
$\delta/2<\alpha<1/2$ we therefore have
\begin{equation} \label{eq:compare-binary-ms}
 F_2(\delta;r_\alpha)\leq B_2^{\mathrm{MS\text{-}CW}}(\delta,\alpha), \qquad
 F_2(\delta;u_2)\leq \min\{\kappa_H(\delta),\kappa_{\mathrm{CW}}(\delta)\}. 
\end{equation}
Indeed, two equally weighted flags carrying $W$ and its
input-swapped copy at swapped compositions have mean $u_2$ and give
\[
 \pe(u_2,\widetilde W)=\pe(r_\alpha,W),\qquad
 \chi_2(u_2,\widetilde W)=1-h_2(\alpha)+\chi_2(r_\alpha,W).
\]
This is precisely the masking operation of
\cite[Definition~2.9 and Propositions~15--16]{AG26}, whose formulas~\cite[(22)--(23)]{AG26} already establish these identities.
In particular, the 2MQC bound of \cite[Corollary~24]{AG26} satisfies
\[
 F_2(\delta;u_2)\leq R_{\mathrm{2MQC}}(\delta).
\]
The masked output is a weighted direct sum of two qubit states: it belongs to
the flagged qMS optimization, and need not itself be a qubit channel.
Thus MQC, 2MQC and the moving-subspace bounds above are recovered by
explicit feasible flag families in $F_2$.

For comparison with Honeycomb, Proposition~8.1 of~\cite{GJL26}
realizes a Horn pair of size $m$ by binary states of rank at most $m$;
Proposition~8.4 identifies its size-two bound with $\kappa_{\mathrm{HC}}$.
The definition of our flagged levels therefore gives
\begin{equation} \label{eq:compare-honeycomb}
 B_2^{(2)}(\delta;u_2)\leq \kappa_{\mathrm{HC}}(\delta). 
\end{equation}
The following comparisons are already proved in
\cite[Theorem~1.2]{GJL26}, for $0<\delta<1/2$:
\begin{equation} \label{eq:known-binary-comparisons}
 \begin{aligned}
 \min\{\kappa_H(\delta),\kappa_{\mathrm{CW}}(\delta)\}
 &\leq R_{\mathrm{2MQC}}(\delta),\\
 \min\{\kappa_{\mathrm{HC}}(\delta),\kappa_{\mathrm{CW}}(\delta)\}
 &<R_{\mathrm{2MQC}}(\delta).
 \end{aligned}
\end{equation}
The second comparison concerns the combined bound; it does not assert
that the Honeycomb branch alone strictly improves 2MQC.

\subsection{The selected envelope and its closure}
\label{sec:selected-envelope}

All six nonbinary comparisons above pass to the closures by
Proposition~\ref{prop:compact-closure} and monotonicity of $\Pp$ and $\Ss$.
For clarity, the envelope selected in~\cite[\S6]{ABCR26} is
\[
\begin{aligned}
 E_q(\delta,\omega)= \min \bigl\{&
 (\Ss U_q^{\BHL,\unr})(\delta), 
 (\Ss U_q^{\MC})(\delta), 
 B_q^{\mathrm{rel\text{-}BE}}(\delta,\omega), \\
 & B_q^{\BHL}(\delta,\omega),
 (\Pp B_q^{\mathrm{cc\text{-}Pl}})(\delta;r_\omega),
 (\Pp B_q^{\mathrm{cc\text{-}CA}})(\delta;r_\omega)
 \bigr\}. 
\end{aligned}
\]
The two unrestricted inputs already include shortening, while the
two constant-composition inputs already include permutation transport.
This selected envelope can subsequently be transported as a whole.
ABCR~\cite[Proposition~6.4, (53)]{ABCR26} also prove that
transport followed by the Bassalygo--Elias reduction gives the same
unrestricted value as that reduction alone.

\begin{corollary}[Comparison with the selected envelope]
\label{cor:envelope}
For $q\geq 3$, $0\leq \omega\leq 1$, and $0<\delta\leq 1$,
\begin{equation}\label{eq:envelope-domination}
 F_q(\delta;r_\omega)\leq E_q(\delta,\omega).
\end{equation}
For $0<\omega<1$ and $0<\delta<\delta_q^{\mathrm{Pl}}(r_\omega)$,
\[
 \UC(\delta;r_\omega)
 <B_q^{(q)}(\delta;r_\omega)
 <F_q(\delta;r_\omega)\leq E_q(\delta,\omega).
\]
\end{corollary}
\begin{proof}
At the native compositions, the six inequalities are
\eqref{eq:compare-mc}, \eqref{eq:compare-bhl-cw},
\eqref{eq:compare-bhl-unr}, \eqref{eq:compare-rel-be},
\eqref{eq:compare-cc-pl}, and~\eqref{eq:compare-cc-ca}.
Apply Proposition~\ref{prop:compact-closure}.
Unrestricted inputs extend from $u_q$ to every $r$ because
$K_q(r,u_q)=0$. Likewise, a weight-only input at $r_\omega$ extends
to every $r$ with $r_0=1-\omega$, since $K_q(r,r_\omega)=0$.
Taking the minimum proves~\eqref{eq:envelope-domination}; the strict
inequalities are Proposition~\ref{prop:compact-strict}.
\end{proof}

\subsection{Comparison summary}
\label{sec:comparison-summary}

Recall that for the unrestricted binary problem, Corollary~\ref{cor:salmon}
and~\eqref{eq:horn-universal-infimum} already give
\[
 \inf_{m\geq 1}B_2^{(m)}(\delta;u_2)
 =B_2^{\mathrm{UC}}(\delta;u_2)=\mathcal J(\delta),
 \qquad 0<\delta<\tfrac12,
\]
where $\mathcal J$ is Salmon's exact binary Delsarte exponent
\cite[Theorem~1.1 and Remark~1.2]{salmon2026rate}.


\begingroup
\definecolor{cmpInk}{RGB}{35,49,65}
\definecolor{cmpBlue}{RGB}{32,83,119}
\definecolor{cmpPale}{RGB}{242,247,250}
\tikzset{
  cmp box/.style={draw=cmpInk!65,rounded corners=2pt,line width=.5pt,
    fill=white,align=center,inner xsep=5pt,inner ysep=5pt,
    font=\sffamily\fontsize{8.5}{10.2}\selectfont,text=cmpInk},
  cmp target/.style={cmp box,fill=cmpPale,draw=cmpBlue},
  cmp narrow/.style={inner xsep=4pt},
  cmp family/.style={cmp box,text width=2.25cm,minimum height=1cm},
  cmp source/.style={cmp box,text width=5.8cm,minimum height=1.15cm},
  cmp result/.style={cmp target,text width=3.3cm,minimum height=.9cm},
  cmp note/.style={font=\sffamily\fontsize{7.3}{8.7}\selectfont,
    text=cmpInk,align=center,fill=white,inner sep=1.5pt},
  cmp above/.style={cmp note,above,yshift=.5mm},
  cmp line/.style={draw=cmpBlue,line width=.85pt},
  cmp arrow/.style={cmp line,-{Stealth[length=5pt,width=4pt]}},
  cmp weak/.style={cmp arrow,densely dashed},
  cmp strict/.style={cmp arrow},
  cmp inclusion/.style={cmp arrow,densely dashed},
  cmp exact/.style={cmp line,-,double=white,double distance=1.3pt},
  cmp heading/.style={font=\sffamily\bfseries\fontsize{9}{11}\selectfont,
    text=cmpInk,anchor=west}
}

\begin{figure}[p]
\centering
\begin{tikzpicture}[x=1cm,y=1cm]
\path (-.45,0);

\begin{scope}
\node[cmp heading] at (0,.55)
  {A. CHANNEL FAMILIES AND RANK HIERARCHY};
\node[cmp family] (mc) at (1.3,-.4)
  {$q$-MC\\[2pt]\cite[\S7, Table~3]{AG26}};
\node[cmp family] (fpc) at (5.2,-.4)
  {FPC-$q$\\[2pt]Prop.~\ref{prop:qmqc}};
\node[cmp family] (qms) at (9.1,-.4)
  {$\mathcal W_q$ (qMS)\\[2pt]Prop.~\ref{prop:qms-seeds}};
\node[cmp family,cmp target] (compact) at (13,-.4)
  {Flagged qMS\\$F_q(\delta;r)$\\\eqref{eq:flagged-qms}};
\draw[cmp inclusion] (mc.east)
  --node[cmp above]{\eqref{eq:mc-embedding}} (fpc.west);
\draw[cmp inclusion] (fpc.east)
  --node[cmp above]{Prop.~\ref{prop:qms-seeds}} (qms.west);
\draw[cmp exact] (qms.east)
  --node[cmp above,above=2pt]{relabel\\and flag\\Lem.~\ref{lem:flags}} (compact.west);

\node[cmp target,text width=3cm,minimum height=.65cm] (fq)
  at (3.1,-1.95) {$F_q(\delta;r)$};
\node[cmp target,text width=3cm,minimum height=.65cm] (bq)
  at (11.3,-1.95) {$B_q^{(q)}(\delta;r)$};
\draw[cmp strict] (fq.east)
  --node[cmp above]{Proposition~\ref{prop:compact-strict}} (bq.west);

\node[cmp target,cmp narrow,text width=2.32cm,minimum height=.75cm] (bm)
  at (1.8,-3.2) {$B_q^{(m)}(\delta;r)$};
\node[cmp target,cmp narrow,text width=2.32cm,minimum height=.75cm] (bmp)
  at (7.5,-3.2) {$B_q^{(m+1)}(\delta;r)$};
\node[cmp target,cmp narrow,text width=2.96cm,minimum height=.75cm] (uc)
  at (13,-3.2) {$B_q^{\mathrm{UC}}(\delta;r)$};
\draw[cmp strict] (bm.east)
  --node[cmp above]{Theorem~\ref{thm:strict}} (bmp.west);
\draw[cmp strict] (bmp.east)
  --node[cmp above]{\eqref{eq:horn-universal-infimum}\\
    Theorem~\ref{thm:strict}} (uc.west);
\end{scope}

\begin{scope}[yshift=-5cm]
\node[cmp heading] at (0,.55)
  {B. SIX SELECTED NONBINARY INPUTS: $q\ge3$};
\node[cmp source] (unr) at (3.1,-.75)
  {\textbf{Unrestricted}\\[2pt]
   $\Ss U_q^{\MC}$\quad\cite[\S7, Table~3]{AG26}\\[2pt]
   $\Ss U_q^{\BHL,\unr}$\quad\cite{BHL07}\\[2pt]
   Shortening: \cite[\S5.6]{ABCR26}};
\node[cmp result] (unrtarget) at (13,-.75)
  {$F_q(\delta;u_q)$\\[2pt]uniform composition};
\draw[cmp weak] (unr.east)
  --node[cmp above]{\eqref{eq:compare-mc}, \eqref{eq:compare-bhl-unr}\\
    Prop.~\ref{prop:compact-closure}} (unrtarget.west);

\node[cmp source] (cw) at (3.1,-2.75)
  {\textbf{Constant weight}\\[2pt]
   $B_q^{\mathrm{rel\text{-}BE}}$\quad\cite[\S5.2, (42)]{ABCR26}\\[2pt]
   $B_q^{\BHL}$\quad\cite{BHL07}; \cite[\S5.4]{ABCR26}};
\node[cmp result] (cwtarget) at (13,-2.75)
  {$F_q(\delta;r_\omega)$\\[2pt]symmetric composition};
\draw[cmp weak] (cw.east)
  --node[cmp above]{\eqref{eq:compare-rel-be}, \eqref{eq:compare-bhl-cw}}
    (cwtarget.west);

\node[cmp source] (cc) at (3.1,-4.45)
  {\textbf{Constant composition}\\[2pt]
   $\Pp B_q^{\mathrm{cc\text{-}Pl}}$,\quad
   $\Pp B_q^{\mathrm{cc\text{-}CA}}$\\[2pt]
   \cite[\S5.8]{ABCR26}};
\node[cmp result] (cctarget) at (13,-4.45)
  {$F_q(\delta;r)$\\[2pt]prescribed composition};
\draw[cmp weak] (cc.east)
  --node[cmp above]{\eqref{eq:compare-cc-pl}, \eqref{eq:compare-cc-ca}\\
    Prop.~\ref{prop:compact-closure}} (cctarget.west);
\end{scope}

\begin{scope}[yshift=-11.25cm]
\node[cmp heading] at (0,.55) {C. BINARY SPECIALIZATIONS};
\node[cmp source] (mqc) at (3.4,-.4)
  {\textbf{MQC and masked MQC}\\[2pt]
   $R_{\mathrm{MQC}}=\kappa_H$\quad\cite[Cor.~23]{AG26}; \cite[\S2]{Bar26}\\[2pt]
   $R_{\mathrm{2MQC}}$\quad\cite[Cor.~24]{AG26}};
\node[cmp result] (f2) at (13,-.4) {$F_2(\delta;u_2)$};
\draw[cmp weak] (mqc.east)
  --node[cmp above]{\cite[Props.~15--16]{AG26}\\
    Lemma~\ref{lem:flags}} (f2.west);

\node[cmp source] (ms) at (3.4,-2)
  {\textbf{Constant-weight moving subspaces}\\[2pt]
   $B_2^{\mathrm{MS\text{-}CW}}(\delta,\alpha)$\\[2pt]
   \cite[Ch.~2, Thm.~3.4, (4)--(7)]{OAI26}};
\node[cmp result] (f2cw) at (13,-2)
  {$F_2(\delta;r_\alpha)$\\[2pt]$\delta/2<\alpha<1/2$};
\draw[cmp weak] (ms.east)
  --node[cmp above]{Proposition~\ref{prop:layer}\\
    \eqref{eq:compare-binary-ms}} (f2cw.west);

\node[cmp box,cmp narrow,text width=3.28cm,minimum height=1.05cm] (hc) at (2.55,-3.7)
  {Honeycomb $\kappa_{\mathrm{HC}}$\\[2pt]
   \cite[Props.~8.1, 8.4]{GJL26}};
\draw[cmp strict] (mqc.west)
  -- (-.15,-.4) -- (-.15,-3.7) -- (hc.west);
\node[cmp note,rotate=90] at (-.07,-2.05)
  {$R_{MQC}$ only\\\cite[Thm.~1.2]{GJL26}};
\node[cmp target,cmp narrow,text width=2.16cm,minimum height=1.05cm] (b2) at (8,-3.7)
  {$B_2^{(2)}(\delta;u_2)$};
\node[cmp target,cmp narrow,text width=3.04cm,minimum height=1.45cm] (j) at (13.1,-3.7)
  {$B_2^{\mathrm{UC}}(\delta;u_2)=\mathcal J(\delta)$\\[2pt]
   {\fontsize{7.3}{8.7}\selectfont
    \cite[Thm.~1.1, Rem.~1.2]{salmon2026rate}\\Cor.~\ref{cor:salmon}}};
\draw[cmp weak] (hc.east)
  --node[cmp above]{\eqref{eq:compare-honeycomb}} (b2.west);
\draw[cmp strict] (b2.east)
  --node[cmp above]{\eqref{eq:horn-universal-infimum}\\
    Thm.~\ref{thm:strict}} (j.west);
\node[cmp box,text width=4.1cm,minimum height=.95cm] (r2) at (2.25,-6.15)
  {$R_{\mathrm{2MQC}}(\delta)$, \cite[Cor.~24]{AG26}};
\node[cmp target,text width=4.8cm,minimum height=.85cm] (weak) at (12.5,-5.4)
  {$\min\{\kappa_H(\delta),\kappa_{\mathrm{CW}}(\delta)\}$, \cite{Bar26,OAI26}};
\node[cmp target,text width=4.8cm,minimum height=.85cm] (strict) at (12.5,-6.95)
  {$\min\{\kappa_{\mathrm{HC}}(\delta),\kappa_{\mathrm{CW}}(\delta)\}$, \cite{GJL26,OAI26}};
\draw[cmp weak] ([yshift=3pt]r2.east) -| (5.,-5.4)
  --node[cmp above]{\cite[Thm.~1.2]{GJL26}} (weak.west);
\draw[cmp strict] ([yshift=-3pt]r2.east) -| (5.,-6.95)
  --node[cmp above]{\cite[Thm.~1.2]{GJL26}} (strict.west);
\end{scope}
\end{tikzpicture}

\caption{\textbf{Channel families, realizations and bound comparisons.}
In the top row of Panel A, dashed arrows denote inclusion of channel families.
The double line indicates the identities from Lemma~\ref{lem:flags}.
For comparisons of bounds $X\to Y$, dashed arrows mean $Y\le X$
and solid arrows mean $Y < X$.
Strict comparisons in panel A assume
$0<\delta<\delta_q^{\mathrm{Pl}}(r)$, with $m\ge1$.
In panel B, $0\le\omega\le1$, and the displayed value of $F_q$ is no larger than either bound in the corresponding source box. 
Zero-cost transport also gives $F_q(\delta;r)\le F_q(\delta;u_q)$ for every $r$, and
$F_q(\delta;r)\le F_q(\delta;r_\omega)$ when $r_0=1-\omega$.
Panel C assumes $0<\delta<1/2$. The side arrow
means $\kappa_{HC} < R_{MQC}$.
The bottom two arrows record
\eqref{eq:known-binary-comparisons}, with strictness applying
to the indicated minimum.}
\label{fig:channel-comparisons}
\end{figure}
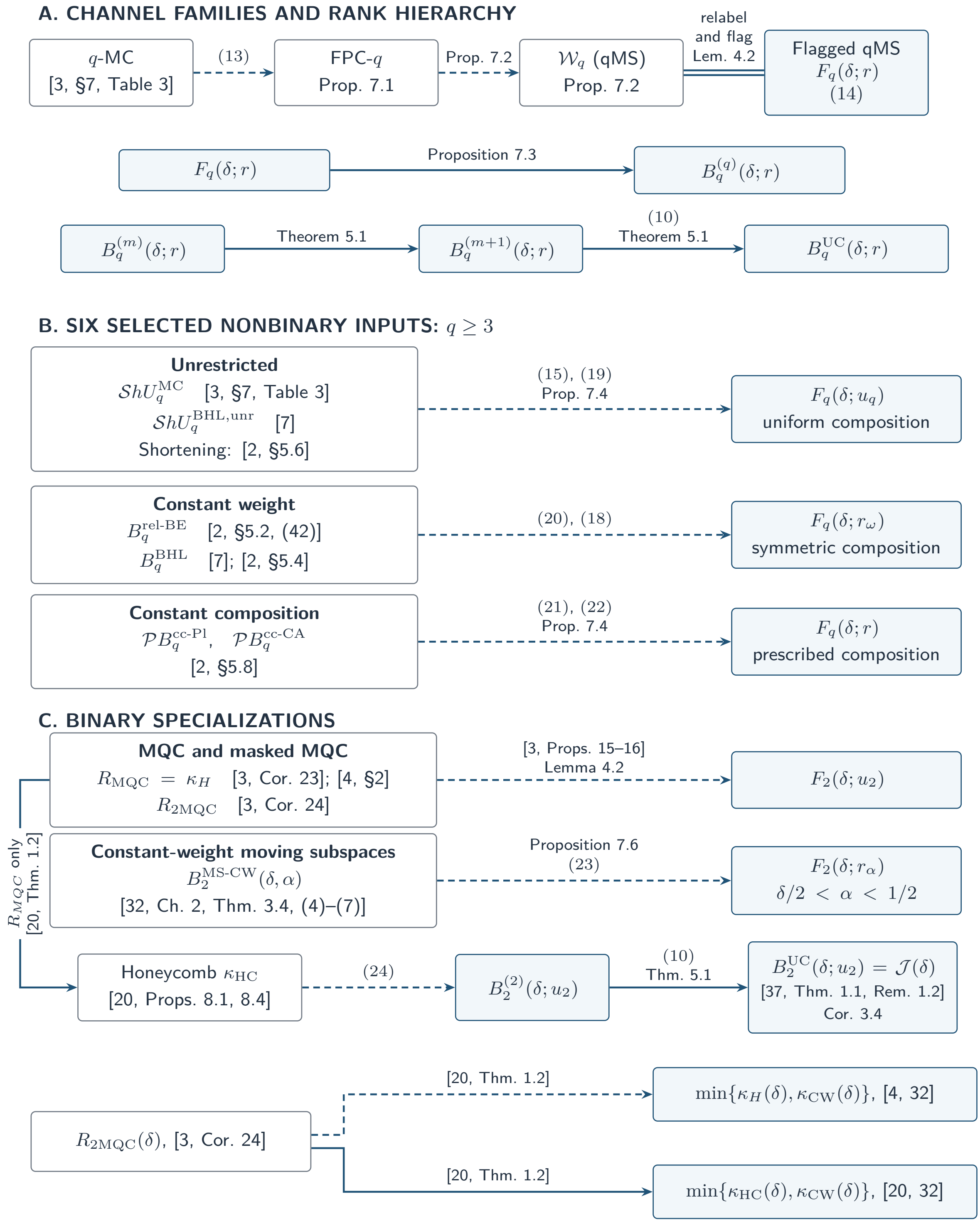
\endgroup

Figure~\ref{fig:channel-comparisons} summarizes the main channel-family inclusions and bound comparisons used in this paper.

\section{Explicit finite matrix formulas}
\label{sec:computation}

This section collects the preceding characterizations of $F_q$ and
$B_q^{(m)}$ in explicit finite matrix form, together with the definitions
needed to evaluate them.

Fix an integer $q\ge2$ and the input alphabet $\A=\{0,\ldots,q-1\}$.
Let $r=(r_x)_{x\in\A}\in\Delta_{q-1}$ be the prescribed composition,
where $\Delta_{q-1}$ is the simplex of probability vectors on $\A$,
and let $0<\delta\le1$ be the relative distance.

We recall from Section~\ref{sec:prelim} the entropy of the composition $r$
and that of a positive semidefinite matrix $\sigma$:
\[
 H_q(r)=-\sum_{x\in\A}r_x\log_qr_x,
 \qquad
 S_q(\sigma)=-\Tr\sigma\log_q\sigma.
\]
We use $0\log0=0$; the second formula applies to positive semidefinite
matrices of any trace.

The argument of Theorem~\ref{thm:main}(b) allows us to use $q+1$ flag
labels $a=1,\ldots,q+1$ in both general searches, with flag probabilities
$w_a$. Sums over input symbols run over $\A$. Conditions indexed by $a$
or $x$ hold for every such index.

\subsection{Flagged qMS}
\label{sec:recipe-qms}

Let $\Perm_q$ denote the permutation group of $\A$. For each flag $a$,
choose output states $\rho_{a,x}\in\mathbb C^{q\times q}$ for $x\in\A$,
an input probability vector $v_a=(v_{a,x})_{x\in\A}\in\Delta_{q-1}$,
and an input permutation $\pi_a\in\Perm_q$.

The states define the channel $W_a:x\mapsto\rho_{a,x}$. The channel
$W_a\circ\pi_a$ assigns $\rho_{a,\pi_a(x)}$ to input $x$, so its average
output state at composition $v_a$ is
\[
 A_a=\sum_xv_{a,x}\rho_{a,\pi_a(x)}.
\]
Its Holevo information and PGM error are
\[
\begin{aligned}
 \chi_q(v_a,W_a\circ\pi_a)
 &=S_q(A_a)-\sum_xv_{a,x}S_q(\rho_{a,\pi_a(x)}),\\
 \pe(v_a,W_a\circ\pi_a)
 &=1-\sum_xv_{a,x}^{\,2}
 \Tr\!\left(
 \rho_{a,\pi_a(x)}A_a^{-1/2}
 \rho_{a,\pi_a(x)}A_a^{-1/2}
 \right).
\end{aligned}
\]
As in Section~\ref{sec:prelim}, $A_a^{-1/2}$ is the inverse square root
on $\supp A_a$ and is zero on its orthogonal complement.

For $g\in\Perm_q$, let $P_g$ be the permutation matrix defined by
$P_ge_x=e_{g(x)}$ on the standard coordinate vectors of $\mathbb C^q$.
Writing $\rho=(\rho_{a,x})$, $v=(v_a)$, $\pi=(\pi_a)$, and $w=(w_a)$,
define the admissible set
\[
\mathcal D_q^{\mathrm{qMS}}(\delta;r)
=
\left\{
(\rho,v,\pi,w):
\begin{array}{l}
\rho_{a,x}\in\mathbb C^{q\times q},\quad
\rho_{a,x}\succeq0,\quad \Tr\rho_{a,x}=1,\\[2pt]
\rho_{a,g(x)}=P_g\rho_{a,x}P_g^*
\quad(g\in\Perm_q,\ g(0)=0),\\[2pt]
v_{a,x}\ge0,\quad \displaystyle\sum_xv_{a,x}=1,\quad
\pi_a\in\Perm_q,\\[2pt]
w_a\ge0,\quad \displaystyle\sum_aw_a=1,\\[2pt]
\displaystyle\sum_aw_av_a=r,\\[2pt]
\displaystyle\sum_aw_a\,\pe(v_a,W_a\circ\pi_a)\le\delta
\end{array}
\right\}.
\]
Then
\begin{equation}\label{eq:qms-finite}
 F_q(\delta;r)
 =H_q(r)+
 \min_{(\rho,v,\pi,w)\in\mathcal D_q^{\mathrm{qMS}}(\delta;r)}
 \sum_aw_a\bigl[\chi_q(v_a,W_a\circ\pi_a)-H_q(v_a)\bigr].
\end{equation}

For each $a$, the matrices $(\rho_{a,x})_{x\in\A}$ satisfying the
covariance and trace conditions are parametrized by at most $13$ real
parameters, as shown in Section~\ref{sec:qms}.

Since $\sum_xv_{a,x}=1$, each component composition requires $q-1$
parameters. Similarly, $\sum_aw_a=1$ determines one of the $q+1$ flag
weights from the other $q$. The total number of real parameters for
$q+1$ flags is therefore at most
\[
 (q+1)\bigl[13+(q-1)\bigr]+q
 =(q+1)(q+13)-1.
\]

\subsection{Flagged Horn-type level \texorpdfstring{$m$}{m}}
\label{sec:recipe-horn}

Fix an integer $m\ge1$. For each flag $a$, choose a positive semidefinite
matrix $Q_a\in\mathbb C^{qm\times qm}$ with $\Tr Q_a^2=1$. Write
\[
 Q_a=(Q_{a,xy})_{x,y\in\A},
 \qquad Q_{a,xy}\in\mathbb C^{m\times m}.
\]

The composition, error and entropy offset defined in
Section~\ref{sec:main} are evaluated by
\[
\begin{aligned}
 \comp(Q_a^2)_x
 &=\Tr\bigl((Q_a^2)_{xx}\bigr),\\
 \pe(Q_a^2)
 &=\sum_x\sum_{y\ne x}\Tr\bigl(Q_{a,xy}^*Q_{a,xy}\bigr),\\
 \phi(Q_a^2)
 &=S_q(Q_a^2)-\sum_xS_q\bigl((Q_a^2)_{xx}\bigr).
\end{aligned}
\]

Writing $Q=(Q_a)$, define
\[
\mathcal D_q^{(m)}(\delta;r)
=
\left\{
(Q,w):
\begin{array}{l}
Q_a\in\mathbb C^{qm\times qm},\quad
Q_a\succeq0,\quad \Tr Q_a^2=1,\\[2pt]
w_a\ge0,\quad \displaystyle\sum_aw_a=1,\\[2pt]
\displaystyle\sum_aw_a\,\comp(Q_a^2)=r,\\[2pt]
\displaystyle\sum_aw_a\,\pe(Q_a^2)\le\delta
\end{array}
\right\}.
\]
Then
\begin{equation}\label{eq:general-gram-search}
 B_q^{(m)}(\delta;r)
 =H_q(r)+
 \min_{(Q,w)\in\mathcal D_q^{(m)}(\delta;r)}
 \sum_aw_a\phi(Q_a^2).
\end{equation}

The matrices $Q_a^2$ range over all Gram data of level $m$ after smaller
slots are padded with zeros.

Each flag has its own Hermitian matrix $Q_a$, specified by $(qm)^2$ real
parameters. The normalization $\Tr Q_a^2=1$, together with $Q_a\succeq0$,
determines one diagonal entry from the remaining entries. Thus the matrix
of each flag requires $(qm)^2-1$ real parameters.

Since $\sum_aw_a=1$, the weights of the $q+1$ flags require another $q$
parameters. The total is therefore
\[
 (q+1)\bigl[(qm)^2-1\bigr]+q
 =(q+1)(qm)^2-1.
\]

\subsection{Unrestricted and constant-weight codes}

Two initial flags suffice for unrestricted codes and three for
constant-weight codes. Only the data of these initial flags need to be
chosen; their copies under the symmetrization of Lemma~\ref{lem:sym},
using cyclic shifts or permutations fixing $0$, respectively, remain
separate flags and introduce no additional parameters.

In the admissible sets, use two labels at $r=u_q$ without prescribing
the average composition of the initial flags, or three labels at
$r=r_\omega$ with only the average zero coordinate prescribed as
$1-\omega$, where $\omega\in[0,1]$ is the relative weight.

With the data at each initial flag fixed, the argument of
Theorem~\ref{thm:main}(b) reduces the number of positive weights $w_a$
while preserving their sum and the average error, together with the
scalar composition condition in the constant-weight case. These two or
three scalar quantities give the stated numbers of initial flags.
All other constraints remain unchanged, and the reduced searches have
the same minima as the general searches.

The counts in Sections~\ref{sec:recipe-qms} and~\ref{sec:recipe-horn}
now apply to two or three initial flags, with one or two weight parameters
after normalization. The resulting upper bounds on the number of real
parameters are:
\begin{center}
\begin{tabular}{lccc}
\toprule
Bound & Unrestricted & Constant weight & Constant composition \\
\midrule
$F_q$ & $2q+25$ & $3q+38$ & $(q+1)(q+13)-1$ \\
$B_q^{(m)}$ & $2(qm)^2-1$ & $3(qm)^2-1$ & $(q+1)(qm)^2-1$ \\
\bottomrule
\end{tabular}
\end{center}

\section{Short rational certificates for finite levels}
\label{sec:certificates}

Every feasible choice in~\eqref{eq:general-gram-search} gives a code-rate
upper bound. We show that nearly optimal matrices and weights admit short rational
descriptions satisfying the composition and error constraints exactly.
These descriptions can be checked independently of the optimizer;
the threshold problem with a prescribed gap consequently belongs to
$\mathsf{PromiseNP}$.

\subsection{Exactly feasible rational descriptions}

Put $D=qm$. We retain the functions $\comp$, $\pe$ and $\phi$ of Section~\ref{sec:main}.
Consider $k$ matrices $Q_1,\ldots,Q_k\in\mathbb C^{D\times D}$,
partitioned into $q$ slots of size $m$, and corresponding weights
$w_1,\ldots,w_k$. All sums over $a$ run from $1$ to $k$.
These matrices and weights are \emph{exactly feasible} at $(\delta;r)$ if
\begin{equation}\label{eq:certificate-feasibility}
 \begin{gathered}
 1\le k\le q+1,\qquad
 Q_a=Q_a^*\succeq0,\qquad Q_a\ne0,\qquad
 G_a=\frac{Q_a^2}{\Tr Q_a^2},\\
 w_a>0,\qquad \sum_aw_a=1,\qquad
 \sum_aw_a\comp(G_a)=r,\qquad \sum_aw_a\pe(G_a)\le\delta.
 \end{gathered}
\end{equation}
Rounding may destroy the normalization $\Tr Q_a^2=1$ used in
Section~\ref{sec:recipe-horn}, so we include normalization
in the definition of $G_a$.
This change of variables does not restrict the feasible set or change
the minimum: every Gram datum is obtained by taking $Q_a=G_a^{1/2}$.
Such matrices and weights certify the upper bound $H_q(r)+\sum_aw_a\phi(G_a)$.

Rationals are encoded by binary numerators and positive binary denominators.
Let $\ell$ be the total bit length of $(q,1^m,r,\delta,\varepsilon)$, with
$q$ in binary, $m$ in unary, and all coordinates of $r$ listed.
Thus $q,m\le\ell$ and $D\le\ell^2$: the input explicitly accounts for
the matrix size.

\begin{theorem}[Short exactly feasible certificates]
\label{thm:rational-certificates}
Let $q\ge2$, $m\ge1$, and let $r\in\mathbb Q^q\cap\Delta_{q-1}$,
$0<\delta\le1$, $0<\varepsilon\le1$ be rational.
There are matrices $\widetilde Q_a$ with rational real and imaginary
entries and rational weights $w_a$ such that, with
\[
 \widetilde G_a
 =\frac{\widetilde Q_a^2}{\Tr\widetilde Q_a^2},
\]
the family $(\widetilde Q_a,w_a)_a$ satisfies
\eqref{eq:certificate-feasibility} and
\begin{equation}\label{eq:certificate-approximation}
 0\le H_q(r)+\sum_aw_a\phi(\widetilde G_a)-B_q^{(m)}(\delta;r)\le\varepsilon.
\end{equation}
Their total description length is $O\bigl((q+1)D^2\ell\bigr)$ bits.
\end{theorem}

\begin{proof}
Choose $Q_a\succeq0$ with $\Tr Q_a^2=1$ and weights $w_a>0$
attaining the minimum in~\eqref{eq:general-gram-search}.
If $r_x=0$, feasibility gives
$\Tr(Q_a^2)_{xx}=\sum_y\|(Q_a)_{xy}\|_F^2=0$, so the corresponding
rows and columns vanish in every matrix. Throughout the proof, perform
all changes only in the blocks with $r_x>0$ and leave the other rows and
columns zero; $q$ and $D$ keep their original values.

Choose $0<h\le1/4$ so that
\[
 12D\sqrt h\le\varepsilon/2,\qquad
 2h\le\varepsilon\delta/2,\qquad
 2h\le\varepsilon r_x/2\quad\text{for every }r_x>0.
\]
These conditions can be satisfied by a rational $h$ whose binary
encoding uses $O(\ell)$ bits.

\emph{1. Round the matrices.}
Choose an integer $b=O(\ell)$ with $2^{-b}\le h/(4D^2)$.
Round the real and imaginary parts of the entries to the nearest
multiples of $2^{-b}$, preserving Hermitian symmetry.
The difference between the rounded matrix and $Q_a$ has Frobenius norm
at most $D2^{-b}$.
Then add $D2^{-b}$ to each diagonal entry in the retained blocks, so that
the resulting matrix $\widetilde Q_a$ is positive semidefinite.
Moreover,
\[
 \|\widetilde Q_a-Q_a\|_F
 \le D2^{-b}+D\sqrt D\,2^{-b}\le2D^2 2^{-b}\le h/2,
\]
which also ensures $\widetilde Q_a\ne0$. Set
\[
 \widehat Q_a=\frac{\widetilde Q_a}{\|\widetilde Q_a\|_F}.
\]
Then $\|\widehat Q_a\|_F=\|Q_a\|_F=1$, and the triangle inequality gives
\begin{equation}\label{eq:certificate-rounding-distance}
 \|\widehat Q_a-Q_a\|_F
 \le |1-\|\widetilde Q_a\|_F|+\|\widetilde Q_a-Q_a\|_F
 \le2\|\widetilde Q_a-Q_a\|_F\le h.
\end{equation}

Fix $a$ and let $R_x,\widehat R_x$ be the block columns of
$Q_a,\widehat Q_a$. Then $\comp(Q_a^2)_x=\|R_x\|_F^2$, and the triangle
inequality and Cauchy--Schwarz give
\begin{equation}\label{eq:certificate-composition-continuity}
\begin{aligned}
 &\sum_x|\comp(Q_a^2)_x-\comp(\widehat Q_a^2)_x| \quad\le\sum_x(\|R_x\|_F+\|\widehat R_x\|_F)\|R_x-\widehat R_x\|_F\\
 &\quad\le\left[\sum_x(\|R_x\|_F+\|\widehat R_x\|_F)^2\right]^{1/2}
 \left[\sum_x\|R_x-\widehat R_x\|_F^2\right]^{1/2}
 \le2h.
\end{aligned}
\end{equation}
The two factors are at most $\|Q_a\|_F+\|\widehat Q_a\|_F=2$ and
$\|Q_a-\widehat Q_a\|_F\le h$, respectively.

For the error, let $A,\widehat A$ be obtained from $Q_a,\widehat Q_a$
by setting their diagonal blocks to zero. Deleting entries can only
decrease the sum of their squared moduli, so
\[
 \|A\|_F\le\|Q_a\|_F=1,\qquad
 \|\widehat A\|_F\le\|\widehat Q_a\|_F=1,\qquad
 \|A-\widehat A\|_F\le\|Q_a-\widehat Q_a\|_F\le h.
\]
Since $\pe(Q_a^2)=\|A\|_F^2$ and
$\pe(\widehat Q_a^2)=\|\widehat A\|_F^2$, it follows that
\begin{equation}\label{eq:certificate-error-continuity}
 |\pe(Q_a^2)-\pe(\widehat Q_a^2)|
 \le(\|A\|_F+\|\widehat A\|_F)\|A-\widehat A\|_F\le2h.
\end{equation}

For entropy, let $0<x<y\le1$ and put $s=y-x$.
The identity $y\ln y-x\ln x=s\ln y+x\ln(1+s/x)$, together with
$\ln(1+u)\le u$ and $y\ge s$, gives
\[
 |y\ln y-x\ln x|
 \le s\ln(1/y)+x\ln(1+s/x)
 \le s\bigl(1+\ln(1/s)\bigr)\le2\sqrt s.
\]
The last inequality follows from $\ln u\le u-1$ with $u=1/\sqrt s$.
By symmetry and continuity at zero, this yields
\begin{equation}\label{eq:certificate-scalar}
 |x\log_qx-y\log_qy|\le\frac{2\sqrt{|x-y|}}{\ln q}
 \qquad(0\le x,y\le1),
\end{equation}
with $0\log_q0=0$.

For the Gram data $Q_a^2$ and $\widehat Q_a^2$, the identity
\[
 Q_a^2-\widehat Q_a^2
 =Q_a(Q_a-\widehat Q_a)+(Q_a-\widehat Q_a)\widehat Q_a
\]
together with~\eqref{eq:certificate-rounding-distance} gives
\[
 \|Q_a^2-\widehat Q_a^2\|_F
 \le(\|Q_a\|_F+\|\widehat Q_a\|_F)\|Q_a-\widehat Q_a\|_F\le2h.
\]
Deleting the off-diagonal blocks gives the same bound for the difference
of $\bigoplus_x(Q_a^2)_{xx}$ and $\bigoplus_x(\widehat Q_a^2)_{xx}$.
For Hermitian matrices $H,K$ of order $D$, with eigenvalues
listed in nondecreasing order, Weyl's eigenvalue perturbation
bound~\cite[Corollary~III.2.6]{Bha97} gives
\[
 |\lambda_j(H)-\lambda_j(K)|
 \le \|H-K\|_{\mathrm{op}}
 \le \|H-K\|_F
 \qquad(j=1,\ldots,D),
\]
where $\|\cdot\|_{\mathrm{op}}$ denotes the operator norm.
Use this with the two pairs
\[
 (H,K)=(Q_a^2,\widehat Q_a^2)
\]
and
\[
 (H,K)=
 \left(\bigoplus_x(Q_a^2)_{xx},
 \bigoplus_x(\widehat Q_a^2)_{xx}\right).
\]
In each case the corresponding eigenvalues differ by at most
$2h$ and belong to $[0,1]$, since the matrices are positive
semidefinite and have trace one.
Using~\eqref{eq:certificate-scalar} with
$x=\lambda_j(H)$ and $y=\lambda_j(K)$ and summing over $j$
bounds each entropy difference by $2D\sqrt{2h}/\ln q$.
Hence
\begin{equation}\label{eq:certificate-continuity}
 |\phi(Q_a^2)-\phi(\widehat Q_a^2)|\le12D\sqrt h\le\varepsilon/2.
\end{equation}
Set
\[
 \widetilde G_a = \widehat Q_a^2 =\frac{\widetilde Q_a^2}{\Tr\widetilde Q_a^2}.
\]
For the mean composition $\widetilde r=\sum_aw_a\comp(\widetilde G_a)$,
averaging~\eqref{eq:certificate-composition-continuity} with the original
weights gives
\begin{equation}\label{eq:certificate-mean-composition}
 \sum_x|\widetilde r_x-r_x|\le2h.
\end{equation}
Also, averaging~\eqref{eq:certificate-error-continuity}
and~\eqref{eq:certificate-continuity}, using feasibility and optimality
of the original matrices and weights, gives
\begin{equation}\label{eq:certificate-rounded-bounds}
\begin{aligned}
 \sum_aw_a\pe(\widetilde G_a)&\le\delta+2h,\\
 H_q(r)+\sum_aw_a\phi(\widetilde G_a)
 &\le B_q^{(m)}(\delta;r)+\varepsilon/2.
\end{aligned}
\end{equation}

\emph{2. Restore exact feasibility.}
For each retained block $x$, let $E_x$ have one unit diagonal entry
in that block and zeros elsewhere. Since $E_x^2=E_x$ and $\Tr E_x=1$,
its Gram datum is $E_x$ itself. Its composition has coordinate $1$ at
$x$ and $0$ elsewhere, and $\pe(E_x)=\phi(E_x)=0$.

For each existing matrix $\widetilde Q_a$, replace its weight $w_a$
by $(1-\varepsilon/2)w_a$. Then add the matrices $E_x$, each with weight
$\gamma_x$. The resulting matrices and weights are:
\[
 \bigl(\widetilde Q_a,(1-\varepsilon/2)w_a\bigr)_{a=1}^k,
 \qquad (E_x,\gamma_x)_{x:r_x>0},
 \qquad \gamma_x=r_x-(1-\varepsilon/2)\widetilde r_x.
\]
By~\eqref{eq:certificate-mean-composition}, $\widetilde r_x\le r_x+2h$.
Together with $2h\le\varepsilon r_x/2$, this gives
\[
 \gamma_x\ge\frac{\varepsilon r_x}{2}-(1-\varepsilon/2)2h
 \ge\frac{\varepsilon^2r_x}{4}>0.
\]
Moreover,
\[
 \sum_x\gamma_x=\sum_xr_x-(1-\varepsilon/2)\sum_x\widetilde r_x
 =\varepsilon/2.
\]
Thus the new weights sum to one. For each retained coordinate $y$,
their mean composition is
\[
 (1-\varepsilon/2)\widetilde r_y+\gamma_y=r_y.
\]

Since $\pe(E_x)=0$, \eqref{eq:certificate-rounded-bounds} and
$2h\le\varepsilon\delta/2$ give the following bound for the mean error:
\[
 (1-\varepsilon/2)\sum_aw_a\pe(\widetilde G_a)
 \le(1-\varepsilon/2)(\delta+2h)\le\delta.
\]
Nonnegativity of Holevo information and the identity in
Lemma~\ref{lem:gram} give
$\phi(G)\ge-H_q(\comp(G))\ge-1$; also $\phi(G)\le0$.
Since $\phi(E_x)=0$, the upper bound for the new matrices and weights
satisfies, by~\eqref{eq:certificate-rounded-bounds},
\[
\begin{aligned}
 H_q(r)+(1-\varepsilon/2)\sum_aw_a\phi(\widetilde G_a)
 &\le H_q(r)+\sum_aw_a\phi(\widetilde G_a)+\varepsilon/2\\
 &\le B_q^{(m)}(\delta;r)+\varepsilon.
\end{aligned}
\]

\emph{3. Choose short rational weights.}
Reindex all matrices from step~2 by $a=1,\ldots,k$, where $k$ now
counts both the rounded matrices and the added $E_x$, so $k\le2q+1$.
Retain the notation $\widetilde Q_a,\widetilde G_a$.
The matrices $\widetilde Q_a$ and $\widetilde G_a$ have rational
entries, whereas the weights from step~2 need not be rational.
The compositions $\comp(\widetilde G_a)$ and PGM errors
$\pe(\widetilde G_a)$ are rational, since
\begin{equation}\label{eq:certificate-data}
 \comp(\widetilde G_a)_x=\frac{\Tr(\widetilde Q_a^2)_{xx}}{\Tr\widetilde Q_a^2},\qquad
 \pe(\widetilde G_a)=\frac{\sum_{x\ne y}\|(\widetilde Q_a)_{xy}\|_F^2}{\Tr\widetilde Q_a^2}.
\end{equation}
To choose rational weights without increasing the upper bound, minimize $\sum_au_a\phi(\widetilde G_a)$ subject to
\[
 \sum_au_a\comp(\widetilde G_a)=r,\qquad
 \sum_au_a\pe(\widetilde G_a)\le\delta,\qquad u_a\ge0.
\]
The weights from step~2 satisfy these constraints, and the composition
equations imply $\sum_au_a=1$, so the feasible set is compact.
Choose a minimizing vertex. Its weights are rational because the
constraints are rational. By the linear dependence argument in the
proof of Theorem~\ref{thm:main}(b), at most $q+1$ weights are positive.
Discard the zero-weight components, reindex the remaining matrices,
and denote their weights by $w_a$.

Each entry of $\widetilde Q_a$ uses $O(\ell)$ bits. At the chosen
vertex, the weights $w_a$ solve a nonsingular system of order at most
$q+1$. Its coefficients from~\eqref{eq:certificate-data} and its
right-hand side from $(r,\delta)$ use $O(\ell)$ bits per entry.
Clearing denominators in each equation gives an integer system
with $O(q\ell)$-bit entries. By Cramer's rule, each weight is a ratio
of two determinants. Each determinant expands into at most $(q+1)!$
signed products of at most $q+1$ entries, so its value uses
$O(q^2\ell)$ bits. Since $D=qm\ge q$, the matrices and weights together
use $O((q+1)D^2\ell)$ bits. Feasibility gives the lower bound
in~\eqref{eq:certificate-approximation}, and the minimizing weights
preserve the upper bound from step~2.
\end{proof}

\subsection{The threshold problem and \texorpdfstring{$\mathsf{PromiseNP}$}{PromiseNP}}

A \emph{promise problem} is a pair of disjoint sets
$L_{\mathrm{yes}},L_{\mathrm{no}}\subseteq\{0,1\}^*$.
It belongs to $\mathsf{PromiseNP}$ if there are a polynomial
$P:\mathbb N\to\mathbb N$ and a deterministic verifier
$V:\{0,1\}^*\times\{0,1\}^*\to\{0,1\}$, running in time polynomial
in $|z|+|y|$, such that
\[
\begin{array}{ll}
 z\in L_{\mathrm{yes}}:
 & V(z,y)=1\text{ for some }y\text{ with }|y|\le P(|z|),\\
 z\in L_{\mathrm{no}}:
 & V(z,y)=0\text{ for every }y\text{ with }|y|\le P(|z|).
\end{array}
\]
Here $y$ is the certificate and $P(|z|)$ bounds its length.
No behavior is required outside the promise.

Consider inputs $(q,1^m,r,\delta,T,\varepsilon)$ with the rational data
of Theorem~\ref{thm:rational-certificates} and $T\in\mathbb Q$, and the promise
\begin{equation}\label{eq:horn-promise}
 \begin{aligned}
 L_{\mathrm{yes}}:&\quad B_q^{(m)}(\delta;r)\le T-\varepsilon,\\
 L_{\mathrm{no}}:&\quad B_q^{(m)}(\delta;r)\ge T+\varepsilon.
 \end{aligned}
\end{equation}

We use the following computational fact. Let $A=A^*\succeq0$ be a
$d\times d$ matrix with rational real and imaginary entries and
$\Tr A=1$, and let $\alpha_1\le\cdots\le\alpha_d$ be its eigenvalues.
For every integer $b\ge1$, a deterministic algorithm computes rational
numbers $\widetilde\alpha_1\le\cdots\le\widetilde\alpha_d$ in $[0,1]$ such that
\[
 |\widetilde\alpha_j-\alpha_j|\le2^{-b}\qquad(j=1,\ldots,d).
\]
Its running time is polynomial in $d$, $b$, and the total number of bits
encoding the entries of $A$.

Compute the characteristic polynomial of $A$ in exact rational arithmetic. Its coefficients have polynomial bit length and can
be computed in polynomial time. Its roots are the eigenvalues of $A$
and lie in $[0,1]$.
Apply the polynomial root approximation result to obtain complex rational
approximations to all these roots with absolute error at most $2^{-b}$
in polynomial time; see \cite[Corollary~2.1.2]{Pan}.
Take the real parts of these approximations, replace values below zero
by zero and values above one by one, and sort the resulting rational numbers.
These operations preserve the required error bound.

\begin{corollary}\label{cor:horn-promise-np}
The promise problem~\eqref{eq:horn-promise} belongs to
$\mathsf{PromiseNP}$.
\end{corollary}
\begin{proof}
A certificate consists of the rational matrices $\widetilde Q_a$ and
weights $w_a$, with the length bound of
Theorem~\ref{thm:rational-certificates} for accuracy $\varepsilon/4$.
Reject malformed or overlong certificates. Check
$\widetilde Q_a=\widetilde Q_a^*\succeq0$, $\widetilde Q_a\ne0$, and
\eqref{eq:certificate-feasibility}, using exact rational arithmetic
and~\eqref{eq:certificate-data}. For each Hermitian $\widetilde Q_a$, test whether $\widetilde Q_a\succeq0$
by checking that all coefficients of the characteristic polynomial of
$-\widetilde Q_a$ are nonnegative. These checks take polynomial time.

For a matrix $A$ as above, apply~\eqref{eq:certificate-scalar} to its
eigenvalues and their rational approximations. This gives
\[
 \left|S_q(A)+\sum_{j=1}^d\widetilde\alpha_j
 \log_q\widetilde\alpha_j\right|
 \le\frac{2d\,2^{-b/2}}{\ln q}.
\]
Choose $b$ so that the right-hand side is at most $\varepsilon/48$;
only $b=O(1+\ln d+\ln(1/\varepsilon))$ is needed.
Evaluate the logarithms so that the sum has additional error at most
$\varepsilon/48$. This takes polynomial time and gives a rational
approximation to $S_q(A)$ with error at most $\varepsilon/24$.
Adding and subtracting $\varepsilon/24$ gives rational lower and upper
bounds differing by $\varepsilon/12$.

Apply this computation to $\diag r$, each $\widetilde G_a$, and each
$\bigoplus_x(\widetilde G_a)_{xx}$. In
$H_q(r)+\sum_aw_a\phi(\widetilde G_a)$, substitute upper bounds for
entropies being added and lower bounds for those being subtracted.
Since $\sum_aw_a=1$, the resulting rational number $U$ satisfies
\[
 H_q(r)+\sum_aw_a\phi(\widetilde G_a)\le U
 \le H_q(r)+\sum_aw_a\phi(\widetilde G_a)+\varepsilon/4.
\]
Accept the certificate exactly when $U\le T$.

On a YES input, Theorem~\ref{thm:rational-certificates} with accuracy
$\varepsilon/4$ supplies a certificate for which
\[
 U\le B_q^{(m)}(\delta;r)+\varepsilon/2
 \le T-\varepsilon/2<T.
\]
On a NO input, every feasible certificate satisfies
\[
 U\ge H_q(r)+\sum_aw_a\phi(\widetilde G_a)
 \ge B_q^{(m)}(\delta;r)\ge T+\varepsilon>T.
\]
Thus every YES input has an accepted certificate of polynomial length,
and no NO input has an accepted certificate.
\end{proof}

\section{Numerical evaluation}
\label{sec:numerics}

The tables compare the flagged qMS bound $F_q(\delta;r)$ and the flagged Horn-type
bound $B_q^{(q)}(\delta;r)$ at level $q$ with the ABCR bounds~\cite{ABCR26}
specified in the captions. The two channel bounds are evaluated
using~\eqref{eq:qms-finite} and~\eqref{eq:general-gram-search}, respectively.
All numerical bounds in the tables are rounded up to seven decimal places. For some parameter choices, a feasible Horn certificate at a level $m<q$
already reproduces the reported level-$q$ value after rounding up to
seven decimal places.

\begin{table}[H]
\centering\small
\setlength{\tabcolsep}{5pt}
\begin{tabular}{ccccc}
\toprule
$q$ & $\delta$ & ABCR~\cite{ABCR26} &
\shortstack{Flagged qMS\\$F_q$} &
\shortstack{Flagged Horn-type level $q$\\$B_q^{(q)}$} \\
\midrule
3 & 0.10 & $0.7763851$ & $0.7763321$ & $0.7763263$ \\
3 & 0.25 & $0.5192018$ & $0.5180825$ & $0.5176789$ \\
3 & 0.30 & $0.4376839$ & $0.4364554$ & $0.4354244$ \\
3 & 0.40 & $0.2770794$ & $0.2769362$ & $0.2766072$ \\
\midrule
4 & 0.10 & $0.8085364$ & $0.8085065$ & $0.8085039$ \\
4 & 0.25 & $0.5830800$ & $0.5823833$ & $0.5821814$ \\
4 & 0.40 & $0.3680736$ & $0.3673653$ & $0.3663631$ \\
\midrule
8 & 0.10 & $0.8519159$ & $0.8519000$ & $0.8518989$ \\
8 & 0.25 & $0.6688171$ & $0.6683584$ & $0.6682190$ \\
8 & 0.40 & $0.4898940$ & $0.4887739$ & $0.4884208$ \\
\bottomrule
\end{tabular}
\caption{Nonbinary unrestricted codes at $r=u_q$: comparison of the ABCR bound with $F_q$ and $B_q^{(q)}$.
The ABCR column is the minimum of the generalized-shortening
closures of the $q$-MC and unrestricted BHL bounds~\cite[\S6]{ABCR26}.}
\label{tab:unrestricted}
\end{table}

\begin{table}[H]
\centering\small
\setlength{\tabcolsep}{5pt}
\begin{tabular}{cccccc}
\toprule
$q$ & $\omega$ & $\delta$ & ABCR~\cite{ABCR26} &
\shortstack{Flagged qMS\\$F_q$} &
\shortstack{Flagged Horn-type level $q$\\$B_q^{(q)}$} \\
\midrule
2 & 0.2 & 0.10 & $0.4144856$ & $0.4144856$ & $0.4144538$ \\
2 & 0.2 & 0.15 & $0.2947205$ & $0.2947205$ & $0.2945334$ \\
2 & 0.2 & 0.25 & $0.0812336$ & $0.0812336$ & $0.0812199$ \\
\bottomrule
\end{tabular}
\caption{Binary constant-weight codes at $q=2$ and $r=r_{0.2}$: comparison of the ABCR bound with $F_q$ and $B_q^{(q)}$.
The ABCR column is obtained by applying permutation transport
to the binary constant-weight moving-subspace bound
described in \cite[\S5.7]{ABCR26}.}
\label{tab:binary-cw}
\end{table}

\begin{table}[H]
\centering\small
\setlength{\tabcolsep}{5pt}
\begin{tabular}{cccccc}
\toprule
$q$ & $\omega$ & $\delta$ & ABCR~\cite{ABCR26} &
\shortstack{Flagged qMS\\$F_q$} &
\shortstack{Flagged Horn-type level $q$\\$B_q^{(q)}$} \\
\midrule
3 & 0.8 & 0.170 & $0.6124319$ & $0.6121265$ & $0.6120643$ \\
3 & 0.8 & 0.175 & $0.6039093$ & $0.6035728$ & $0.6035015$ \\
3 & 0.8 & 0.180 & $0.5954212$ & $0.5950514$ & $0.5949701$ \\
\midrule
3 & 0.3 & 0.20 & $0.3468652$ & $0.3463384$ & $0.3462042$ \\
\midrule
4 & 0.3 & 0.15 & $0.4087064$ & $0.4085865$ & $0.4085694$ \\
4 & 0.3 & 0.20 & $0.3340410$ & $0.3337173$ & $0.3336497$ \\
4 & 0.3 & 0.25 & $0.2614698$ & $0.2607731$ & $0.2605712$ \\
\bottomrule
\end{tabular}
\caption{Nonbinary constant-weight codes at $r=r_\omega$: comparison of the ABCR bound with $F_q$ and $B_q^{(q)}$.
The ABCR column is obtained by applying permutation transport
to the constant-weight BHL bound described in \cite[\S5.4]{ABCR26}.}
\label{tab:cw}
\end{table}

The most pronounced relative improvements over the ABCR bounds occur at
compositions $r$ that are not permutations of any $r_\omega$.

\begin{table}[H]
\centering\small
\setlength{\tabcolsep}{5pt}
\begin{tabular}{cccccc}
\toprule
$q$ & $r$ & $\delta$ & ABCR~\cite{ABCR26} &
\shortstack{Flagged qMS $F_q$} &
\shortstack{Flagged Horn-type level $q$ $B_q^{(q)}$} \\
\midrule
3 & $(0.65,0.30,0.05)$ & 0.45 & $0.0357248$ & $0.0115303$ & $0.0115291$ \\
3 & $(0.65,0.30,0.05)$ & 0.46 & $0.0254056$ & $0.0064459$ & $0.0064450$ \\
3 & $(0.65,0.30,0.05)$ & 0.47 & $0.0151775$ & $0.0026270$ & $0.0026264$ \\
\bottomrule
\end{tabular}
\caption{Constant-composition codes at $q=3$ and $r=(0.65,0.30,0.05)$: comparison of the ABCR bound with $F_q$ and $B_q^{(q)}$.
The ABCR column is obtained by applying permutation transport
to the constant-composition Plotkin bound described in \cite[\S5.8]{ABCR26}.}
\label{tab:cc}
\end{table}

\begin{table}[H]
\centering\small
\setlength{\tabcolsep}{5pt}
\begin{tabular}{cccccc}
\toprule
$q$ & $r$ & $\delta$ & ABCR~\cite{ABCR26} &
\shortstack{Flagged qMS\\$F_q$} &
\shortstack{Flagged Horn-type level $q$\\$B_q^{(q)}$} \\
\midrule
4 & $(0.45,0.45,0.05,0.05)$ & 0.51 & $0.0664177$ & $0.0348756$ & $0.0326367$ \\
4 & $(0.45,0.45,0.05,0.05)$ & 0.52 & $0.0578444$ & $0.0279062$ & $0.0260918$ \\
4 & $(0.45,0.45,0.05,0.05)$ & 0.53 & $0.0493538$ & $0.0215253$ & $0.0201091$ \\
\bottomrule
\end{tabular}
\caption{Constant-composition codes at $q=4$ and $r=(0.45,0.45,0.05,0.05)$: comparison of the ABCR bound with $F_q$ and $B_q^{(q)}$.
The ABCR column is obtained by applying permutation transport
to the constant-composition Plotkin bound described in \cite[\S5.8]{ABCR26}.}
\label{tab:cc-631}
\end{table}

\appendix

\section{Pure-state channels near a classical channel}
\label{app:rays}

Let $\psi$ denote the pure-state channel $x\mapsto\psi_x\psi_x^*$, where
$\psi_x=\Theta^{1/2}e_x$, the vectors $e_x$ form the standard
basis, and $\Theta\succeq0$ has unit diagonal. For a composition $v\in\Delta_{q-1}$, put
$D=\diag v$ and $\Gamma=D^{1/2}\Theta D^{1/2}$.
Lemma~\ref{lem:gram} with $R_x=\sqrt{v_x}\psi_x$ gives
\begin{equation}\label{eq:pure}
 \chi_q(v,\psi)=S_q(\Gamma),\qquad 1-\pe(v,\psi)=\sum_x\bigl[(\Gamma^{1/2})_{xx}\bigr]^2,
\end{equation}
and every $\Gamma\succeq0$ with diagonal $v$ arises in this way. Let
$\sqrt v$ denote the entrywise square root, $Q=I-\sqrt v\sqrt v^{\mathsf T}$ and
$M(v)=QDQ$; then $\Tr M(v)=\delta^\mathrm{Pl}_q(v)$, and $M(v)\succ0$ on
$\sqrt v^\perp$ when $v$ has full support and $q\ge2$.

\begin{lemma}\label{lem:sqrt-perturb}
Let $w$ be a real unit vector, $Q=I-ww^{\mathsf T}$, $A$ be a real symmetric matrix with $K=QAQ\succ0$ on
$w^\perp$, and $\Gamma(t)=ww^{\mathsf T}+tA\succeq0$ for small $t>0$. Then, as
$t\downarrow0$,
\[
 \Gamma(t)^{1/2}=ww^{\mathsf T}+\sqrt t\,K^{1/2}+O(t),\qquad
 \mathrm{spec}\,\Gamma(t)=\{1+O(t)\}\cup\{t\kappa_j+O(t^2)\},
\]
where $\kappa_j$ are the eigenvalues of $K$ on $w^\perp$.
\end{lemma}

\begin{proof}
The eigenvalue of $\Gamma(t)$ near $1$ is isolated, with gap $1-O(t)$; by the
Davis--Kahan theorem~\cite{DK70} its spectral projection is
$P(t)=ww^{\mathsf T}+O(t)$. Choose a unitary $U=I+X$ with $X=O(t)$ and
$Uww^{\mathsf T}U^*=P(t)$ (for instance the polar part of
$Pww^{\mathsf T}+(I-P)Q$). Then
$U^*\Gamma U=\lambda_0ww^{\mathsf T}\oplus QU^*\Gamma UQ$ with $\lambda_0=1+O(t)$,
and since $Q\Gamma=tQA=O(t)$, $QU^*\Gamma UQ=Q\Gamma Q+O(t^2)=tK+O(t^2)$.
On $w^\perp$, write
\[
 B(t)=\left.QU^*\Gamma(t)UQ\right|_{w^\perp}
 =tK|_{w^\perp}+E(t),
 \qquad \|E(t)\|_{\mathrm{op}}=O(t^2).
\]
Weyl's eigenvalue perturbation bound \cite[Corollary~III.2.6]{Bha97} gives
\[
 |\lambda_j(B(t))-t\kappa_j| \le \|E(t)\|_{\mathrm{op}}=O(t^2).
\]
Together with $\lambda_0=1+O(t)$, this proves the asserted
spectral expansion. 

On $w^\perp$, we have $K\succeq\kappa I$ for some $\kappa>0$.
Hence, for sufficiently small $t$,
\[
 B(t)^{1/2}
 =\sqrt t\,(K+O(t))^{1/2}
 =\sqrt t\,K^{1/2}+O(t^{3/2}),
\]
by~\cite[Theorem~X.3.8]{Bha97}.
Since $\lambda_0^{1/2}=1+O(t)$, conjugating the block
decomposition back by $U=I+O(t)$ gives
\[
 \Gamma(t)^{1/2}
 =ww^{\mathsf T}+\sqrt t\,K^{1/2}+O(t).
\]
\end{proof}

In the asymptotic formulas below, $\pe(t)$ and $\chi_q(t)$
denote the error and Holevo information of the indicated
channel at the specified composition. 

\begin{corollary}[Equiangular ray]\label{cor:rays}
Restricting to the support of $v$, assume that $v$ has full support; and let $\Theta(t)=(1-t)J+tI$, where
$J$ is the all-ones matrix. For the pure-state channel with Gram matrix
$\Theta(t)$ at the composition $v$, as $t\downarrow0$:
\[
 \delta^\mathrm{Pl}_q(v)-\pe(t)=2\sqrt t\,\Tr M(v)^{3/2}+O(t),\qquad
 \chi_q(t)=\delta^\mathrm{Pl}_q(v)\,t\log_q(1/t)+O(t).
\]
\end{corollary}

\begin{proof}
Here $\Gamma(t)=ww^{\mathsf T}+tA$ with $w=\sqrt v$ and $A=D-ww^{\mathsf T}$, so
$\diag A=0$ and $K=QAQ=M(v)$. By Lemma~\ref{lem:sqrt-perturb},
$(\Gamma^{1/2})_{xx}=v_x+\sqrt t\,(K^{1/2})_{xx}+O(t)$, so by
\eqref{eq:pure} $1-\pe=\sum_xv_x^2+2\sqrt t\,\Tr(DK^{1/2})+O(t)$, and
$\Tr(DK^{1/2})=\Tr(QDQK^{1/2})=\Tr M(v)^{3/2}$ since $K^{1/2}=QK^{1/2}Q$. For
$\chi_q=S_q(\Gamma)$ use the spectrum: the top eigenvalue contributes $O(t)$
and each eigenvalue $t\kappa_j+O(t^2)$ contributes
$\kappa_jt\log_q(1/t)+O(t)$, with $\sum_j\kappa_j=\Tr M(v)=\delta^\mathrm{Pl}_q(v)$.
\end{proof}

These channels are the $q$-ary pure-state channels of~\cite[\S7]{AG26}
($\eta=0$ in the $q$-MC) with $t=(\sqrt{1-p}-\sqrt{p/(q-1)})^2$, used at the
composition $v$.

\subsection{Behaviour at the Plotkin boundary}
\label{subsec:plotkin-boundary}

The constant-composition form of the Alrabiah--Guruswami channel criterion
\cite{AG26}, established in Theorem~\ref{thm:main}, also controls the
vanishing of the exact rate at the Plotkin boundary $\delta^{\mathrm{Pl}} := \delta_q^{\mathrm{Pl}}(r)$. We obtain an explicit
composition-dependent coefficient by choosing the perturbation in
Lemma~\ref{lem:sqrt-perturb}. Throughout this subsection, the composition rate
$r$ is fixed: the blocklength tends to infinity first, and only then does
the gap to the Plotkin threshold tend to zero.

Let $k=|\supp r|\ge2$. All matrices below are indexed by
$\supp r$, so that their diagonal weights are positive. Put
\[
 w=\sqrt r,\qquad D= \diag r,\qquad
 Q=I-ww^{\mathsf T},\qquad M=QDQ,
\]
and define
\[
 V(r):=\Tr M^2
 =\sum_xr_x^2-2\sum_xr_x^3+
 \left(\sum_xr_x^2\right)^2.
\]
Since $M$ is positive definite on $w^\perp$, we have $V(r)>0$;
also $\Tr M=\delta^{\mathrm{Pl}}$.

\begin{proposition}[Rate near the Plotkin boundary]
\label{prop:plotkin-boundary}
For every fixed composition rate $r$ as above, there exist positive constants
$\kappa_r,\lambda_r,\varepsilon_r$ such that,
for $0<\varepsilon<\varepsilon_r$,
\[
 \frac{\varepsilon^2}{2V(r)\ln q}-\kappa_r\varepsilon^3 \leq R_q(\delta^{\mathrm{Pl}}-\varepsilon;r) \leq \UC(\delta^{\mathrm{Pl}}-\varepsilon;r)
 \leq \frac{\varepsilon^2}{2V(r)}\log_q\frac1\varepsilon
 +\lambda_r\varepsilon^2.
\]
Consequently,
\begin{equation}\label{eq:plotkin-exponent}
 \lim_{\varepsilon\downarrow0}
 \frac{\ln R_q(\delta^{\mathrm{Pl}}-\varepsilon;r)}{\ln\varepsilon}=2,
 \qquad
 \left.\frac{\partial}{\partial\delta}R_q(\delta;r)
 \right|_{\delta=\delta^{\mathrm{Pl}}}=0.
\end{equation}
The upper bound uses a single pure-state channel with output dimension
$k$.
\end{proposition}

\begin{proof}
Write $V=V(r)$.

\emph{Upper bound.}
Set
\[
 K=M^2,\qquad z_x=\frac{K_{xx}}{2w_x},\qquad
 A=K-wz^{\mathsf T}-zw^{\mathsf T},\qquad
 \Gamma(t)=ww^{\mathsf T}+tA.
\]
Then $\diag A=0$, $QAQ=K$, and $w^{\mathsf T}Aw=-V$.
With $b=Qz$, the block form relative to
$\mathbb Rw\oplus w^\perp$ is
\[
 \Gamma(t)=
 \begin{pmatrix}
 1-tV&-tb^{\mathsf T}\\
 -tb&tK
 \end{pmatrix}.
\]
Here $K$ is positive definite on $w^\perp$. Its Schur complement shows
that $\Gamma(t)$ is positive definite whenever
$0<t<(V+b^{\mathsf T}K^{-1}b)^{-1}$, with the inverse taken on $w^\perp$.
Moreover, $\diag \Gamma(t)=r$ and $\Tr\Gamma(t)=1$.
Thus $\Gamma(t)$ is the weighted Gram matrix of a pure-state
channel, as in~\eqref{eq:pure}; symbols outside the support may be assigned
arbitrary pure output states.

Lemma~\ref{lem:sqrt-perturb} and~\eqref{eq:pure} give
\begin{align*}
 \delta^{\mathrm{Pl}}-\pe(t)
 &=2\sqrt t\,\Tr(D\sqrt K)+O_r(t)
 =2V\sqrt t+O_r(t),\\
 \chi_q(t)
 &=\Tr K\,t\log_q(1/t)+O_r(t)
 =Vt\log_q(1/t)+O_r(t).
\end{align*}
Indeed, $\sqrt K=M$ and $\Tr(DM)=\Tr M^2=V$.
Choose
\[
 \sqrt t=\frac{\varepsilon}{2V}+\eta_r\varepsilon^2
\]
with a sufficiently large fixed $\eta_r>0$. For all sufficiently small
$\varepsilon$, this ensures the strict constraint
$\pe(t)<\delta^{\mathrm{Pl}}-\varepsilon$, without requiring monotonicity of $\pe(t)$.
Since $t=\varepsilon^2/(4V^2)+O_r(\varepsilon^3)$, substitution into
$\chi_q(t)$ proves the desired upper bound.

\emph{Lower bound.}
The constant-composition Varshamov--Gilbert bound
\cite[\S5.9]{ABCR26} can be written as
\begin{equation}\label{eq:plotkin-gv}
 R_q(\delta^{\mathrm{Pl}}-\varepsilon;r)\geq J_r(\varepsilon):=
 \min_{\substack{P\ge0,\ P\mathbf1=P^{\mathsf T}\mathbf1=r\\
 \Tr P\ge\sum_xr_x^2+\varepsilon}}
 D_q(P\Vert rr^{\mathsf T}),
\end{equation}
where
$D_q(P\Vert rr^{\mathsf T})=
\sum_{x,y}P_{xy}\log_q(P_{xy}/(r_xr_y))$.
For completeness, we compute its local expansion. Put $P_0=rr^{\mathsf T}$.
The matrix
\[
 P_*=P_0+\frac{\varepsilon}{V}D^{1/2}MD^{1/2}
\]
is feasible for small $\varepsilon$: the perturbation has zero row and
column sums and trace $\varepsilon$. Taylor expansion at the strictly
positive matrix $P_0$ gives
\[
 D_q(P_*\Vert P_0)
 =\frac{\varepsilon^2}{2V\ln q}+O_r(\varepsilon^3).
\]
Let $P_\varepsilon$ be a minimizer in~\eqref{eq:plotkin-gv}. The preceding
trial matrix and Pinsker's inequality imply
$\|P_\varepsilon-P_0\|_1=O_r(\varepsilon)$, where this is the entrywise
$\ell^1$ norm. Set
$B=D^{-1/2}(P_\varepsilon-P_0)D^{-1/2}$.
The marginal constraints give $B=QBQ$, and
\[
 \langle M,B\rangle_F=\Tr(P_\varepsilon-P_0)\ge\varepsilon.
\]
Here $\langle X,Y\rangle_F=\Tr(X^{\mathsf T}Y)$ is the Frobenius inner
product. A uniform Taylor expansion and Cauchy--Schwarz now yield
\[
 J_r(\varepsilon)
 =\frac{\|B\|_F^2}{2\ln q}+O_r(\varepsilon^3)
 \ge\frac{\varepsilon^2}{2V\ln q}+O_r(\varepsilon^3).
\]
Together with the trial matrix, this proves
\[
 J_r(\varepsilon) =\frac{\varepsilon^2}{2V(r)\ln q}+O_r(\varepsilon^3),
\]
which implies the desired lower bound.
The type-class greedy argument also gives~\eqref{eq:plotkin-gv} for
irrational $r$ under our definition of the rate: use types converging to
$r$ on the same support, the polynomial number of joint types, and
continuity of entropy.

The bounds imply
$R_q(\delta^{\mathrm{Pl}}-\varepsilon;r)=\varepsilon^{2+o(1)}$
and $R_q(\delta^{\mathrm{Pl}}-\varepsilon;r)=o(\varepsilon)$.
Since Theorem~\ref{thm:main} gives $R_q(\delta;r)=0$ for
$\delta\ge\delta^{\mathrm{Pl}}$, both conclusions in~\eqref{eq:plotkin-exponent} follow.
\end{proof}

\paragraph{Optimality within affine pure-state perturbations.}
For any affine ray $\Gamma(t)=ww^{\mathsf T}+tA$ with $A$ real symmetric,
$\diag A=0$, and $K=QAQ$ positive definite on $w^\perp$,
the same argument gives the leading coefficient
\[
 \frac{\Tr K}{2[\Tr(M\sqrt K)]^2}
 \ge\frac1{2\Tr M^2}=\frac1{2V(r)}.
\]
Equality holds precisely when $\sqrt K$ is a positive scalar multiple of
$M$. Thus the choice $K=M^2$ is optimal within this class of rays.
This statement does not assert optimality over all pure-state paths,
flagged channels, or the full optimization defining $\UC$.
For comparison, the equiangular ray in Corollary~\ref{cor:rays} gives
$\delta^{\mathrm{Pl}}/(2[\Tr M^{3/2}]^2)$, which is no smaller because
$[\Tr M^{3/2}]^2\leq \Tr M\,\Tr M^2$.
At $r=u_q$, one has $V(r)=(q-1)/q^2$, recovering the coefficients in the
classical Varshamov--Gilbert and first LP expansions
\cite{MRRW77,AG26}; the pure-state realization belongs to the
Alrabiah--Guruswami framework~\cite{AG26}.

\paragraph{Classical comparison and the order of limits.}
The logarithmic exponent and zero derivative in
\eqref{eq:plotkin-exponent} also follow from classical bounds.
For a word of composition rate $r$, the centered embedding
\[
 v(x)=(e_{x_i}-r)_{i=1}^n
\]
satisfies $\|v(x)\|^2=n\delta^{\mathrm{Pl}}$ and
$\langle v(x),v(y)\rangle=n\delta^{\mathrm{Pl}}-d_H(x,y)$, in dimension at most
$(k-1)n$. The spherical Kabatiansky--Levenshtein bound~\cite{KabLev78}
(see also~\cite[Eq.~(2.6)]{CZ14}), expanded at maximal inner product
$\varepsilon/\delta^{\mathrm{Pl}}$, therefore gives
\[
 R_q(\delta^{\mathrm{Pl}}-\varepsilon;r)
 \leq \frac{k-1}{2(\delta^{\mathrm{Pl}})^2}\varepsilon^2\log_q\frac1\varepsilon
 +O_r(\varepsilon^2).
\]
The argument extends to converging types; symbols outside
$\supp r$ can first be fixed and deleted at an $o(n)$ loss
in the logarithm of the code size. Since
$V(r)\ge(\delta^{\mathrm{Pl}})^2/(k-1)$, the coefficient in
the upper bound of Proposition~\ref{prop:plotkin-boundary} is at least as strong as this direct
spherical comparison. This comparison alone does not establish an
improvement over all known constant-composition LP bounds.

There is also a substantial literature on shrinking neighbourhoods of the
Plotkin threshold. Krasikov and Litsyn~\cite{KrLi97} combine linear
programming with a Plotkin--Johnson argument for binary codes with
$d=(n-j)/2$ and $0<j<n^{1/3}$. Boyvalenkov, Danev, and
Stoyanova~\cite{BDS18} use the discreteness of Hamming distances to refine
Levenshtein's bounds; their explicit low-degree asymptotics include
polynomial-size regimes with $(1-1/q)n-d=O(\sqrt n)$.
Pang, Mahdavifar, and Pradhan~\cite{PMP23} obtain constructions and
Fourier-analytic LP bounds in the large-minimum-distance regime, including
polynomial upper bounds for $A_2(n,\lceil n/2-\rho\sqrt n\rceil)$
for specified fixed values of $\rho$.
These results concern a gap that shrinks with the blocklength.
In contrast, Proposition~\ref{prop:plotkin-boundary} first takes the
asymptotic rate at fixed $\varepsilon>0$ and then lets
$\varepsilon\downarrow0$. Without a uniform finite-length remainder, it
does not justify substituting $\varepsilon=\varepsilon_n$ or deducing
polynomial bounds on code cardinality in those shrinking-gap regimes.

\section{Symbol merging}
\label{app:merging}

Let $2\leq q'\leq q$, let $\varphi:[q]\to[q']$ be a symbol map, and put
\[
 Q_y=\sum_{x:\varphi(x)=y}r_x,
\]
where $r\in\Delta_{q-1}$. For each class $c_y=\varphi^{-1}(y)$ with $Q_y>0$, write
$r_y^{\max}=\max_{x\in c_y}r_x$ and define
\[
 \lambda=\sum_{y:Q_y>0}\min\{Q_y,\,2(Q_y-r_y^{\max})\},\qquad
 \mu=\sum_{y:Q_y>0}\left(Q_y-\frac{\sum_{x\in c_y}r_x^2}{Q_y}\right),
 \qquad a=\log_q q'.
\]
Zero-mass classes are omitted throughout.
Here $\lambda$ is the maximal relative distance loss under coordinatewise
merging of two words of composition rate $r$: within a class, at most $nQ_y$
coordinates can differ, and at most $2n(Q_y-r_y^{\max})$ can involve a
symbol other than one of maximal frequency. Thus the merging argument of
\cite[\S6]{ABCR26}, with any valid smaller-alphabet bound $L$, gives
\[
 R_q(\delta;r)\leq aL(\delta-\lambda;Q),\qquad \delta>\lambda.
\]
The factor $a$ converts rates in base $q'$ to base $q$.

\begin{proposition}[Channel merging sandwich]\label{prop:merging}
For every $\delta>0$,
\[
 \UC(\delta;r)\geq aB^{\mathrm{UC}}_{q'}(\delta;Q).
\]
Moreover $0\leq \mu\leq \lambda$, and for $\delta>\mu$,
\[
 \UC(\delta;r)\leq aB^{\mathrm{UC}}_{q'}(\delta-\mu;Q).
\]
Consequently, for $\delta>\lambda$,
\begin{equation}\label{eq:merging-sandwich}
 aB^{\mathrm{UC}}_{q'}(\delta;Q)
 \leq \UC(\delta;r)
 \leq aB^{\mathrm{UC}}_{q'}(\delta-\mu;Q)
 \leq aB^{\mathrm{UC}}_{q'}(\delta-\lambda;Q).
\end{equation}
\end{proposition}

\begin{proof}
Merge a channel $W:x\mapsto\rho_x$ into
$W':y\mapsto\rho'_y=Q_y^{-1}\sum_{x\in c_y}r_x\rho_x$.
Lemma~\ref{lem:coarse} gives $\pe(Q,W')\leq \pe(r,W)$, while the Holevo
chain rule gives
\[
 \chi_q(r,W)=\chi_q(Q,W')+
 \sum_yQ_y\chi_q(r(\cdot\mid c_y),W|_{c_y})
 \geq a\chi_{q'}(Q,W').
\]
Taking the infimum over channels with error below $\delta$ proves the
lower bound.

Conversely, lift a channel $W':y\mapsto\rho'_y$ by setting
$\rho_x=\rho'_{\varphi(x)}$. The average states agree, so
$\chi_q(r,W)=a\chi_{q'}(Q,W')$. Its PGM operators satisfy
$M_x=(r_x/Q_y)M'_y$ for $x\in c_y$. With
$s_y=\Tr(\rho'_yM'_y)\in[0,1]$, this gives
\[
 \pe(r,W)=\pe(Q,W')+
 \sum_y\left(Q_y-\frac{\sum_{x\in c_y}r_x^2}{Q_y}\right)s_y
 \leq \pe(Q,W')+\mu.
\]
Taking the infimum over channels with $\pe(Q,W')<\delta-\mu$ proves the
upper bound. Each summand of $\mu$ is nonnegative and at most both $Q_y$
and $2(Q_y-r_y^{\max})$, since
$\sum_{x\in c_y}r_x^2\ge(r_y^{\max})^2$.
Thus $\mu\leq \lambda$; the final inequality follows from monotonicity in
the distance.
\end{proof}

If $L(\delta-\lambda;Q)\ge
B^{\mathrm{UC}}_{q'}(\delta-\lambda;Q)$, then
\eqref{eq:merging-sandwich} shows that
$aL(\delta-\lambda;Q)\ge\UC(\delta;r)$.
Symbol merging can therefore transfer an improvement already present on
the smaller alphabet, but cannot create one.

\section{Fixed points of generalized shortening}
\label{app:shortening-fixed-points}

We give an elementary characterization of the fixed points of the
generalized-shortening operator~\eqref{eq:shortening}, based on
the shortening inequality of Laihonen and Litsyn~\cite{LL98}.
The characterization concerns functions; it does not require that
they arise from channels or even be upper bounds on code rates.

Fix $q\ge2$, put $\theta_q=(q-1)/q$, and define
\[
 c_q(z)=1-h_q\!\left(\min\left\{\frac z2,\theta_q\right\}\right), \qquad 0\leq z\leq 2.
\]
Thus $c_q$ is convex and nonincreasing,
$c_q(0)=1$, and $c_q(z)=0$ for $z\ge2\theta_q$.
The substitution $z=2t$ rewrites~\eqref{eq:shortening} as
\[
 (\Ss U)(\delta) = \inf \bigl\{(1-\alpha)U(\gamma)+\alpha c_q(z)\bigr\},
 \qquad 0\leq \delta\leq 1,
\]
where infimum is taken over $0\leq \alpha\leq 1$, $0\leq \gamma\leq 1$, $0\leq z\leq 2$, $\delta=(1-\alpha)\gamma+\alpha z$.
In particular, the auxiliary coordinate $z$ ranges over $[0,2]$,
although the argument of $U$ ranges over $[0,1]$.
We call $U$ \emph{stable under generalized shortening} if $\Ss U=U$.

\begin{proposition}[Idempotence]
\label{prop:shortening-idempotence}
For every bounded real-valued function $U$ on $[0,1]$,
\[
 \Ss U\leq U,\qquad \Ss(\Ss U)=\Ss U.
\]
Moreover, $\Ss U$ is the greatest bounded shortening-stable function
that is pointwise at most $U$.
\end{proposition}

\begin{proof}
The choice $\alpha=0$ gives $\Ss U\leq U$.
The operator is order-preserving, so $\Ss(\Ss U)\leq \Ss U$.
For the converse inequality, consider two successive operations,
\[
 \delta=\alpha z+(1-\alpha)f,
 \qquad f=\beta y+(1-\beta)e.
\]
Put $\gamma=\alpha+(1-\alpha)\beta$. If $\gamma>0$, let
\[
 \bar z=\frac{\alpha z+(1-\alpha)\beta y}{\gamma}\in[0,2].
\]
Then $\delta=\gamma\bar z+(1-\gamma)e$, and convexity of $c_q$ gives
\[
 \alpha c_q(z)+(1-\alpha)
 \bigl[\beta c_q(y)+(1-\beta)U(e)\bigr] \geq \gamma c_q(\bar z)+(1-\gamma)U(e)
 \ge(\Ss U)(\delta).
\]
If $\gamma=0$, the same conclusion follows from $U(\delta)\ge\Ss U(\delta)$.
Taking the two infima proves idempotence.
Finally, if $V\leq U$ and $\Ss V=V$, then
$V=\Ss V\leq \Ss U$, proving maximality.
\end{proof}

\begin{theorem}[Characterization of shortening-stable functions]
\label{thm:shortening-fixed-points}
Let $U:[0,1]\to[0,1]$. The following are equivalent:
\begin{enumerate}[label=(\roman*),leftmargin=2em]
\item $\Ss U=U$.
\item $U(x)\leq c_q(x)$ for $0\leq x\leq 1$, and, for every
$z\in[0,2]$, the function
\begin{equation}\label{eq:shortening-secants}
 x\longmapsto\frac{U(x)-c_q(z)}{x-z}
\end{equation}
is nondecreasing separately on
$[0,1]\cap(-\infty,z)$ and $[0,1]\cap(z,\infty)$.
\end{enumerate}
Every such fixed point is nonincreasing and locally Lipschitz
on $(0,1)$.
If $U$ is continuous on $[0,1]$, these conditions are also
equivalent to:
\begin{enumerate}[label=(\roman*),leftmargin=2em,start=3]
\item $U$ is locally absolutely continuous on $(0,1)$ and,
for almost every $0<\delta<1$,
\begin{equation}\label{eq:shortening-differential}
 U'(\delta)\leq 0,\qquad
 U(\delta)-\delta U'(\delta)
 \leq 1-\log_q\!\left(1+(q-1)q^{2U'(\delta)}\right).
\end{equation}
\end{enumerate}
At each differentiability point of a fixed point, the inequality
in~\eqref{eq:shortening-differential} holds.
\end{theorem}

\begin{proof}
\emph{Secant characterization.}
Since $\Ss U\leq U$, condition (i) is equivalent to
\begin{equation}\label{eq:shortening-all-chords}
 U\bigl((1-\alpha)\gamma+\alpha z\bigr)
 \leq (1-\alpha)U(\gamma)+\alpha c_q(z)
\end{equation}
whenever $\gamma\in[0,1]$, $z\in[0,2]$, $0\leq \alpha\leq 1$, and
the argument on the left belongs to $[0,1]$.
Taking $\alpha=1$ gives $U\leq c_q$ on $[0,1]$.
For $x<y<z$, apply~\eqref{eq:shortening-all-chords} with
$\gamma=x$ and $(1-\alpha)x+\alpha z=y$.
After division by the negative number $y-z$, the result is
\[
 \frac{U(x)-c_q(z)}{x-z}
 \leq \frac{U(y)-c_q(z)}{y-z}.
\]
For $z<x<y$, apply the same inequality with $\gamma=y$ and
$(1-\alpha)y+\alpha z=x$; division by $x-z>0$ gives
the same comparison. This proves (ii).
Conversely, a point strictly between $\gamma$ and $z$ lies on the
same side of $z$ as $\gamma$, and the appropriate monotonicity in (ii)
gives~\eqref{eq:shortening-all-chords} by reversing these calculations.
The remaining cases follow from $U\leq c_q$ or are identities.

\emph{Monotonicity and regularity.}
Use $z=2\theta_q\ge1$ and $c_q(z)=0$ in
\eqref{eq:shortening-all-chords}. For $0\leq x<y<1$ this gives
\[
 U(y)\leq \frac{2\theta_q-y}{2\theta_q-x}U(x)\leq U(x).
\]
The endpoint $y=1$ follows by the same argument; when $q=2$,
use $U(1)\leq c_q(1)=0$.
Next take $z=0$, $\gamma=y$ and $\alpha=1-x/y$.
For $0<x<y\leq 1$ we obtain
\[
 U(x)\leq 1-\frac{x}{y}+\frac{x}{y}U(y),
 \qquad
 0\leq U(x)-U(y)
 \leq \frac{y-x}{x}\bigl(1-U(x)\bigr)
 \leq \frac{y-x}{x}.
\]
This proves local Lipschitz continuity.

\emph{Differential characterization.}
At a differentiability point $\delta$, set $\gamma=\delta$ in
\eqref{eq:shortening-all-chords}, divide by $\alpha>0$, and
let $\alpha\downarrow0$. For every $z\in[0,2]$ we get
\begin{equation}\label{eq:shortening-tangent}
 U(\delta)+(z-\delta)U'(\delta)\leq c_q(z).
\end{equation}
For $z=\delta$, this is simply $U(\delta)\leq c_q(\delta)$.
Thus every tangent line lies below the whole curve $c_q$.
For $p\leq 0$, direct minimization gives
\begin{equation}\label{eq:shortening-legendre}
 \inf_{0\leq z\leq 2}\{c_q(z)-pz\}
 =1-\log_q\!\left(1+(q-1)q^{2p}\right).
\end{equation}
Indeed, on $0\leq z\leq 2\theta_q$ put $t=z/2$.
The maximum of $h_q(t)+2pt$ is attained at
\[
 t=\frac{(q-1)q^{2p}}{1+(q-1)q^{2p}}\leq \theta_q
\]
and equals $\log_q(1+(q-1)q^{2p})$.
On $2\theta_q\leq z\leq 2$, the expression $c_q(z)-pz=-pz$
is minimized at $z=2\theta_q$, already included in the preceding
interval. Equations~\eqref{eq:shortening-tangent} and
\eqref{eq:shortening-legendre} prove (iii).

Conversely, suppose $U$ is continuous on $[0,1]$ and satisfies (iii).
Equation~\eqref{eq:shortening-legendre} implies
\eqref{eq:shortening-tangent} for almost every $\delta$ and every $z$.
For each fixed $z$, the quotient in~\eqref{eq:shortening-secants}
is locally absolutely continuous away from $z$, with derivative
\[
 \frac{(x-z)U'(x)-U(x)+c_q(z)}{(x-z)^2}\ge0
 \quad\text{for almost every }x.
\]
It is therefore nondecreasing on each side of $z$;
continuity extends these comparisons to the endpoints.
Taking $z=x$ in the almost-everywhere tangent inequality gives
$U(x)\leq c_q(x)$ almost everywhere, and continuity gives it
everywhere. Hence (ii) holds.
\end{proof}

If $U$ is convex as well as shortening-stable, the slope inequality
holds at every $0<\delta<1$ for every
$p\in[U'_-(\delta),U'_+(\delta)]$.
Indeed, for $z>\delta$ the directional version of
\eqref{eq:shortening-tangent} uses $U'_+(\delta)$, whereas for
$z<\delta$ it uses $U'_-(\delta)$.
Every intermediate $p$ satisfies both inequalities, and
\eqref{eq:shortening-legendre} applies.

\begin{remark}[A sufficient convexity criterion]
\label{rem:shortening-convexity}
If $U:[0,1]\to[0,1]$ is convex and nonincreasing,
$U(1)=0$, and $U(x)\leq c_q(x)$ on $[0,1]$, then $\Ss U=U$.
For $z\leq 1$, this follows from convexity and $U(z)\leq c_q(z)$.
For $z>1$ and parameters satisfying $\delta=(1-\alpha)\gamma+\alpha z\leq1$, we have
\[
 U(\delta)
 \leq U\bigl((1-\alpha)\gamma+\alpha\bigr)
 \leq (1-\alpha)U(\gamma)+\alpha U(1)
 \leq (1-\alpha)U(\gamma)+\alpha c_q(z),
\]
where the last inequality uses $U(1)=0\leq c_q(z)$.
All channel bounds used in Corollary~\ref{cor:shortening-shape}
satisfy this endpoint condition.
The exact characterization in Theorem~\ref{thm:shortening-fixed-points}
does not assume convexity.
\end{remark}

\paragraph{AI usage.} This work is a result of a study of recent papers~\cite{AG26,GJL26,OAI26} and investigation into the relationship between the new estimates and the transport operator introduced in the recent paper of the authors~\cite{ABCR26}. We used ChatGPT 6.1 Sol and ChatGPT Astra in these studies and for the preparation of the manuscript. The final check was performed by Claude Opus 5.5.
All mathematical statements, proofs, computations, and citations were independently verified by the authors, who take full responsibility for the content of the manuscript.

\paragraph{Acknowledgments.} DC is supported by the Bulgarian National Science Fund Contract KP-06-N72/6-2023.
PB is supported by the Ministry of Education and Science of Bulgaria under the Grant No. DO1-98/26.06.2025 ``National Centre for High Performance and Distributed Computing'' (NCHPDC).

\bibliographystyle{plain}
\bibliography{refs}

\end{document}